\documentclass[letterpaper]{article} 
\usepackage{aaai25}  
\usepackage{times}  
\usepackage{helvet}  
\usepackage{courier}  
\usepackage[hyphens]{url}  
\usepackage{graphicx} 
\usepackage{natbib}  
\usepackage{caption} 
\usepackage{algorithm}

\usepackage{newfloat}
\usepackage{listings}
\DeclareCaptionStyle{ruled}{labelfont=normalfont,labelsep=colon,strut=off} 
\floatstyle{ruled}
\newfloat{listing}{tb}{lst}{}
\floatname{listing}{Listing}
\title{Certification of Speaker Recognition Models to Additive Perturbations}
\author{
    Dmitrii Korzh\textsuperscript{\rm 1,2},
    Elvir Karimov\textsuperscript{\rm 1,2},
    Mikhail Pautov\textsuperscript{\rm 1,4},
    Oleg Y. Rogov\textsuperscript{\rm 1,2,3},
    Ivan Oseledets\textsuperscript{\rm 1,2}
}
\affiliations{
    \textsuperscript{\rm 1}AIRI, Moscow, Russia\\
    \textsuperscript{\rm 2}Skolkovo Institute of Science and Technology, Moscow, Russia\\
    \textsuperscript{\rm 3}Moscow Technical University of Communications and Informatics, Moscow, Russia\\
    \textsuperscript{\rm 4}ISP RAS Research Center for Trusted Artificial Intelligence, Moscow, Russia\\
    korzh@airi.net, karimov@airi.net, pautov@airi.net,
   rogov@airi.net,
    oseledets@airi.net
}

\usepackage{amsmath}
\usepackage{latexsym}
\usepackage{amssymb}
\usepackage{amsthm}
\usepackage{booktabs}
\usepackage{enumitem}
\usepackage{microtype}
\usepackage{tabularx}
\usepackage{mathtools}
\usepackage{subcaption}
\usepackage[noend]{algpseudocode}

\theoremstyle{plain}
\newtheorem{theorem}{Theorem}

\theoremstyle{definition}

\theoremstyle{remark}
\newtheorem{remark}[theorem]{Remark}

\DeclareMathOperator*{\argmin}{argmin}
\DeclareMathOperator*{\argmax}{argmax}
\usepackage{xcolor}

\begin{document}

\maketitle

\begin{abstract}
Speaker recognition technology is applied to various tasks, from personal virtual assistants to secure access systems. However, the robustness of these systems against adversarial attacks, particularly to additive perturbations, remains a significant challenge. In this paper, we pioneer applying robustness certification techniques to speaker recognition, initially developed for the image domain. Our work covers this gap by transferring and improving randomized smoothing certification techniques against norm-bounded additive perturbations for classification and few-shot learning tasks to speaker recognition. We demonstrate the effectiveness of these methods on VoxCeleb 1 and 2 datasets for several models. We expect this work to improve the robustness of voice biometrics and accelerate the research of certification methods in the audio domain.
\end{abstract}

%
\begin{links}
    \link{Code}{https://github.com/AIRI-Institute/asi-certification}
    \link{Extended version}{https://arxiv.org/abs/2404.18791}
\end{links}

\section{Introduction}

This work addresses the issues of robustness and privacy in deep learning voice biometrics models \cite{snyder2018x, wan2018generalized}. Although deep learning models excel in various applications, they are unreliable and susceptible to specific perturbations. These perturbations may be imperceptible to humans but can dramatically affect the model's performance \cite{szegedy2013intriguing, kaviani2022adversarial}. Researchers have developed various methods to compute adversarial perturbations and defenses against them, recently becoming necessary to provide provable guarantees on model behavior under constrained perturbations \cite{li2023sok, cohen2019certified}. However, the audio domain has not received as much attention as the image domain. Given the escalating levels of speech fraud due to advancements in adversarial models and deepfake technologies \cite{qin2023openvoice}, significant security risks could arise in biometric systems or even in creating personalized scams in social networks. Thus, this article focuses on the certification of automatic speaker recognition models. The certified speaker recognition model is the one in which prediction does not change under additive perturbations of the input audio.

Automatic speaker recognition models \cite{desplanques2020ecapa, bredin2020pyannote, wang2023wespeaker} typically utilize spectrograms (such as Mel spectrograms) or raw-waveform frontends to address several vital tasks. The first task is automatic speaker identification (ASI), where the model determines the speaker's identity in an audio recording. The second task is automatic speaker verification (ASV), which involves verifying whether two audio samples are from the same speaker. The third task is speaker diarization, where the model segments audio into parts corresponding to different speakers.

Voice biometric models convert speech into vector representations, ensuring that utterances from the same speaker generate closely aligned vectors while those from different speakers are widely separated. These properties should hold even for speakers not encountered during training. Several training strategies exist for the encoder that maps audio $x$ to these embeddings. One approach uses metric learning with triplet \cite{hermans2017defense} or contrastive loss \cite{wang2021understanding}. Another strategy involves training an embedder combined with a classifier on a fixed set of speakers, with variations of cross-entropy loss that was initially developed for face biometrics \cite{meng2021magface} to enhance the expressiveness and separation of embeddings, even for unseen speakers. During inference, cosine similarity, cosine distance, or other distance metrics are used to match the embedding of the inference audio to the closest reference speaker's embedding (enrollment vector).

\begin{figure*}[htb]
     \centering
        \includegraphics[width=0.77\textwidth]{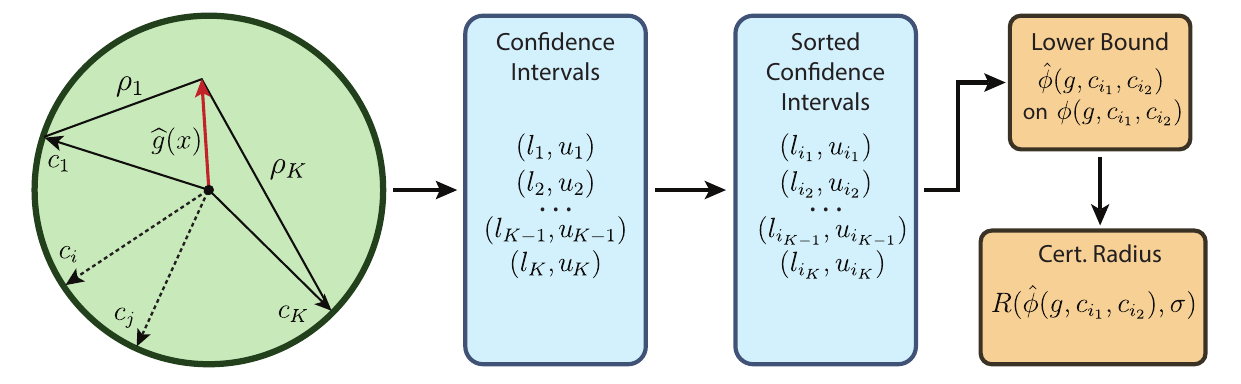}
        \caption{The scheme illustrating the proposed algorithm. The algorithm requires an audio sample $x$, base model $f$, and the set of centroids $S^c=\{c_1, \dots, c_K\}$. In the Figure, $\hat g(x)$ corresponds to the estimation of the smoothed embedding $g(x)$ from  Eq. \eqref{eq:rs_model_smooth} computed in the form from Eq. \eqref{eq:g_mc}. When executed, Algorithm \ref{alg:cr} computes the confidence interval $(l_i, u_i)$ for the distance between $\hat{g}(x)$ and corresponding centroid $c_i$ for all $i \in [1,\dots,K]$. Then, given sorted confidence intervals $\{(l_{i_1}, u_{i_1}), \dots, (l_{i_K}, u_{i_K})\}$, two closest centroids, $c_{i_1}$ and $c_{i_2}$, are determined. The last step of the algorithm is the computation of the lower bound $R(\hat{\phi}(\hat{g},c_{i_1}, c_{i_2}))$ on the certified radius $R({\phi}(g,c_{i_1}, c_{i_2}))$ from the Theorem \ref{th:robustness_guarantee}.}
        \label{fig:main_fig}
\end{figure*}

Our work explores the certified robustness of speaker recognition models against any additive perturbation constrained by the $l_2$ norm value. Such perturbations can be created via various adversarial attacks, whether targeted or untargeted, white-box or black-box scenarios, in which the attacker may know the model's architecture, parameters, and gradients or may only have input and output access. 

\textbf{Our contributions can be summarized as follows:}
\begin{itemize}
    \item We introduce a novel randomized smoothing-based approach to certify few-shot embedding models against additive, norm-bounded perturbations. Our approach provides state-of-the-art certification results in a few-shot setting.
    \item We derive robustness certificates and demonstrate their advantages over those obtained using existing competitors' methods. Our theoretical claims are supported by experimental results on the VoxCeleb datasets using several well-known speaker recognition models.
    \item To the best of our knowledge, there are no previous works that present the provable robustness of speaker recognition models. We highlight this issue and provide starting baselines that others can improve in future research.
\end{itemize}

\section{Related Work}
\subsection{Speaker Recognition}
Recently, speaker recognition \cite{snyder2018x, wan2018generalized, desplanques2020ecapa, wang2023cam++, wang2023wespeaker} has made significant progress. The x-vector system, based on Time Delay Neural Network (TDNN) technology, has been particularly influential and further developed in many other models. This system uses one-dimensional convolution to pick up important time-related features in speech. For example, ECAPA-TDNN \cite{desplanques2020ecapa} uses techniques that allow the model to consider a wider range of time-related information, combining features recursively from several previous states for the next hidden state. Later, a densely connected TDNN (D-TDNN) \cite{yu2020densely} was presented, which reduced the number of parameters needed. Additionally, the Context-Aware Masking (CAM) module, a type of pooling,  was combined with D-TDNN, and the model CAM++ improves the performance regarding verification metrics (such as Equal Error Rate and Detection Cost Function) and the inference time.

\subsection{Adversarial Attacks}
It has long been known \cite{szegedy2013intriguing, goodfellow2014explaining} that deep learning models are vulnerable to small additive perturbations of input. In recent years, many approaches have been proposed to generate adversarial examples, for example, \cite{athalye2018obfuscated, khrulkov2018art, su2019one, Yuan_2021_ICCV, wang2023adversarial}. These methods expose different conceptual vulnerabilities of models: some generate attacks using information about the model's gradient, while others deploy separate networks to produce malicious input. Moreover, adversarial examples can be transferred across models \cite{inkawhich2019feature}, which limits the application of neural networks in various practical scenarios. This vulnerability poses significant risks in contexts such as biomedical image segmentation \cite{apostolidis2021survey}, industrial face recognition \cite{komkov2021advhat} and detection \cite{kaziakhmedov2019real}, self-driving car systems \cite{deng2020analysis}, and speaker recognition systems \cite{zhang2023imperceptible, lan2022adversarial, li2020practical}. Additionally, speaker anonymization systems aim to conceal identity features while preserving other information (text, emotions) from the speech, and often based on the generation of additive perturbations \cite{deng2023v, liu2024transferable}. 

\subsection{Empirical and Certified Defenses}
Numerous defensive approaches have recently been proposed \cite{li2023sok, fan2023robustness} to mitigate the effects of the attacks. Among these, adversarial training \cite{goodfellow2014explaining, andriushchenko2020understanding} is arguably the best technique to enhance the robustness of models in practice. The method is straightforward~--~during training, each batch of data is augmented with adversarial examples generated by a specific method. Consequently, the model becomes more resistant to the type of attack used during the training process. However, the model may easily become overfitted to the provided attacks and unable to be robust against new types of adversarial perturbations. Additionally, adversarial training is time-consuming and often leads to notable performance degradation.
 Despite this, several prominent fast adversarial training approaches exist. Data augmentation with ordinary transforms and noises (e.g., Gaussian) and regularization techniques (e.g., consistency loss \cite{jeong2020consistency}) are the most straightforward, cheapest, and prominent approaches to increase empirical robustness. Additionally, \cite{castan2017improving, zhou2023adversarial, wu2021improving} are improved empirical guarantees in speaker recognition using unlabeled data, adversarial training, and self-supervised methods.

Another research direction is the development of methods that provide provable certificates on the model's prediction under certain transformations. Mainly, approaches are based on Satisfiability Modulo Theory \cite{pulina2010abstraction} and Mixed Integer Linear Programming \cite{cheng2017maximum} solvers, on the interval \cite{gowal2019scalable} and polyhedra \cite{lyu2020fastened} relaxation, on analysis of Lipschitz continuity  \cite{salman2019provably} and the curvature of the decision boundary of the network \cite{singla2020second}. 

Nowadays, randomized smoothing \cite{cohen2019certified, salman2019provably} forms the basis for many certification approaches, offering defenses against both norm-bounded \cite{yang2020randomized} and semantic perturbations \cite{li2021tss, muravevcertified, hao2022gsmooth}. This method is simple, effective, and scalable to large models and datasets. Notably, it can also be theoretically applied to certify automatic speech recognition systems \cite{olivier-raj-2021-sequential}.

\section{Methodology}


In this section, we define the problem statement, provide an overview of the techniques used, and describe the proposed method for certifying embedding models against norm-bounded additive perturbations.

\subsection{Speaker Recognition as a Few-Shot Problem}
Few-shot learning is a machine learning paradigm where models are trained to generalize effectively from only a few examples of each class, addressing the challenge of limited data availability  \cite{koch2015siamese, snell2017prototypical}, that is highly relevant to biometrics systems. Consider $f: \mathbb{R}^n \to \mathbb{R}^d$ as the base model that maps input audios to normalized embeddings, where $\|f(\cdot)\|_2 =1$, $n$ is an input dimension, $d$ is an embedding dimension. After training the embedding model, we need to enroll new speakers we want to authorize later in our biometrics system. 

For every enrolled speaker, the enrollment vector or centroid is established as the mean or weighted sum of embeddings derived from collected audio samples of the speaker. These centroids create the basis for calculating the similarity with the embeddings of new audio samples during inference authorization. The enrollment dataset, denoted as $S^e = \{(x_1, y_1), \dots, (x_l, y_l)\}$, consists of audio samples $x_i \in \mathbb{R}^n$ assigned to corresponding speakers $y_i \in [1,\dots,K]$. Depending on the application, this dataset may consist of speakers not encountered during training or a mix of seen and unseen speakers. For a given class $k$, the subset $S^e_k = \{(x_i, y_i) \in S^e : y_i = k\}$ comprises the audios belonging to the speaker $k$. Although in practice, the number of available audios $M(k)$ in every subset $S^e_k$ can vary from speaker to speaker, in the few-shot setting, the number $M(k)$ is fixed to the pre-defined number $M$ of audios used to construct the speaker's enrollment vector   $\forall k \mapsto |S^e_k|=M$ for the fair comparison. The normalized speaker enrollment embedding (speaker centroid, prototype) is then can be formalized as follows:
\begin{equation}\label{eq:class_prot}
c_k = \frac{1}{M} \sum_{x \in S^e_k} f(x), \ ~ \|c_k\|_2 = 1,
\end{equation}
and a database $S^c = \{c_j\}_{j=1}^{j=K}$ of centroid vectors is constructed.
During inference, a new sample $x \in S^i$ is classified by assigning it to the speaker whose enrollment vector from $S^c$ is the closest in terms of some  distance  function $\rho$:
\begin{equation}\label{eq:fewshotclass}
i_1 = \argmin_{k \in [1,\dots,K]} \rho(f(x), c_k).
\end{equation}
Although few-shot usually implies $M \in [1, 2, 3]$ only,  we consider $M \in [1, 5, 10] $ following common biometrics practice. We equate the speaker recognition (ASI) and few-shot models to emphasize that our method is also applicable to other few-shot scenarios.

\subsection{Problem Statement and Certification for Vector Functions} \label{rs_section}
Certification guarantees against additive perturbations of a bounded magnitude can be formulated as follows. Suppose that $f$ is the base vector (embedding) model,  $c_k$ is defined as in Eq. \eqref{eq:class_prot}, and $R>0$ is the norm threshold. Then, the model $f$ is said to be certified at $x$, if for all $\|\delta\|_2 \le R$,
\begin{equation}
    \argmin_{k \in [1,\dots K]} \rho (f(x), c_k) =  \argmin_{k \in [1,\dots K]} \rho(f(x+\delta), c_k).
\end{equation}
Unfortunately, this cannot be achieved directly for the $f$, but $f$ can be substituted with \emph{smoothed model} $g$. This technique is called a \emph{randomized smoothing (RS)}, and it was initially proposed for the classification \cite{lecuyer2019certified, cohen2019certified} and $g$ has an important property of Lipschitz continuity \cite{salman2019provably}: outputs' perturbation can be limited for the fixed input's perturbation level. Given the classifier model $f_{\text{clf}}: \mathbb{R}^n \to [0,1]^K$ and the smoothing distribution $\mathcal{N}(0,\sigma^2I)$ the smoothed model takes the form 
\begin{equation}\label{eq:rs_model}
    g_{\text{clf}}(x) = \mathbb{E}_{\varepsilon \sim \mathcal{N}(0,\sigma^2I)} f_{\text{clf}}(x+\varepsilon),
\end{equation}
here $g_{\text{clf}}(x)$ is the vector of class probabilities with $K$ components.
As it is shown in \cite{cohen2019certified}, when the model from Eq. \eqref{eq:rs_model} is confident in predicting the correct class $i_1$ for the input $x$, 
\begin{equation}\label{eq:rs_pred}
    g_{\text{clf}}(x)_{i_1} = p_{i_1} \ge p_{i_2} = \max_{i \ne i_1} g_{\text{clf}}(x)_i
\end{equation}
then it is robust in $l_2-$ball around $x$ of radius 
\begin{equation}\label{eq:cert_r}
    R = \frac{\sigma}{2} \left(\Phi^{-1}(p_{i_1}) - \Phi^{-1}(p_{i_2})\right),
\end{equation}
\begin{equation}
    \forall \delta: \|\delta\|_2 < R \mapsto
 \argmax g_{\text{clf}}(x) = \argmax g_{\text{clf}}(x+\delta),
\end{equation}
where $\Phi^{-1}(\cdot)$ is the inverse of the standard Gaussian cumulative density function.

For the vector functions, let us consider the base model $f: \mathbb{R}^n \to \mathbb{R}^d$ that maps input to normalized embeddings,  the associated smoothed model $g: \mathbb{R}^n \to \mathbb{R}^d$ is defined as 
\begin{equation}\label{eq:rs_model_smooth}
    g(x) = \mathbb{E}_{\varepsilon \sim \mathcal{N}(0,\sigma^2I)} f(x+\varepsilon).
\end{equation}
Here $g(x)$ is $d-$dimensional smoothed embedding. Note that $f$ and centroids $c_k$ are normalized while $g$ is not. Suppose that input audio $x$ is correctly assigned to class $i_1$ represented by centroid $c_{i_1}$. Assume that  $c_{i_2}$ is the second closest to $g(x)$ centroid. If we introduce scalar mapping $\phi: \mathbb{R}^d \to [0,1]$ in the form
\begin{equation}\label{eq:phi}
    \phi = \phi(g(x), c_{i_1}, c_{i_2}) = \frac{\langle g(x), c_{i_1}-c_{i_2}\rangle}{2 \|c_{i_1}-c_{i_2}\|_2} + \frac{1}{2},
\end{equation}
then the following robustness guarantee holds:
\begin{theorem}[Main result]\label{th:robustness_guarantee}
For all additive perturbations $\delta: \|\delta\|_2 \le R(\phi, \sigma) = \sigma  \Phi^{-1} (\phi)$
\begin{equation}\label{eq:rob_guarantee}
        \argmin_{k \in [1,\dots K]} \|g(x) - c_k\|_2 =  \argmin_{k \in [1,\dots K]} \|g(x+\delta) - c_k\|_2,
\end{equation}
where $R(\phi,\sigma)$ is called certified radius of $g$ at $x.$
\end{theorem}
\setcounter{theorem}{0}
\begin{remark}
The detailed proof is provided in the Appendix of the full manuscript version.
\end{remark}
\begin{remark}\label{remark2}
     The method is generalizable to open setups and other neural embedding tasks, requiring only the two closest centroids for certification. Thus, it cannot be applied to ASV certification. Note that cosine distance is as suitable as $l_2$ norm. 
\end{remark}


\section{Implementation Details}

\begin{figure*}[htb]
     \centering
     \begin{subfigure}{0.31\textwidth}
         \centering
         \includegraphics[width=\textwidth]{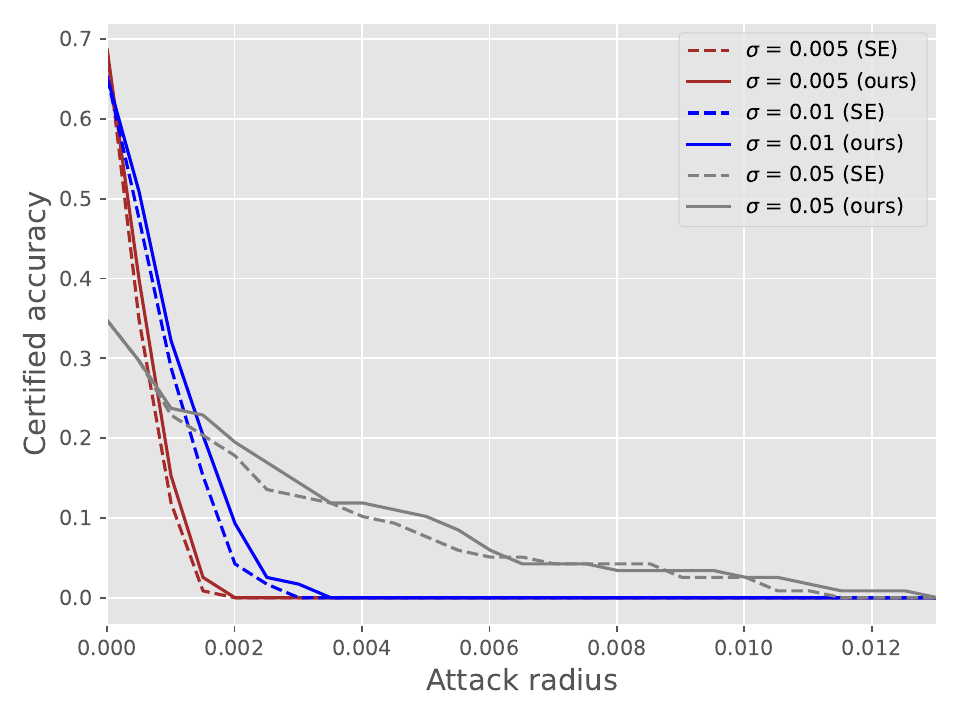}
         \caption{Dependency on $\sigma$.}
     \end{subfigure}
     \begin{subfigure}{0.31\textwidth}
         \centering
         \includegraphics[width=\textwidth]{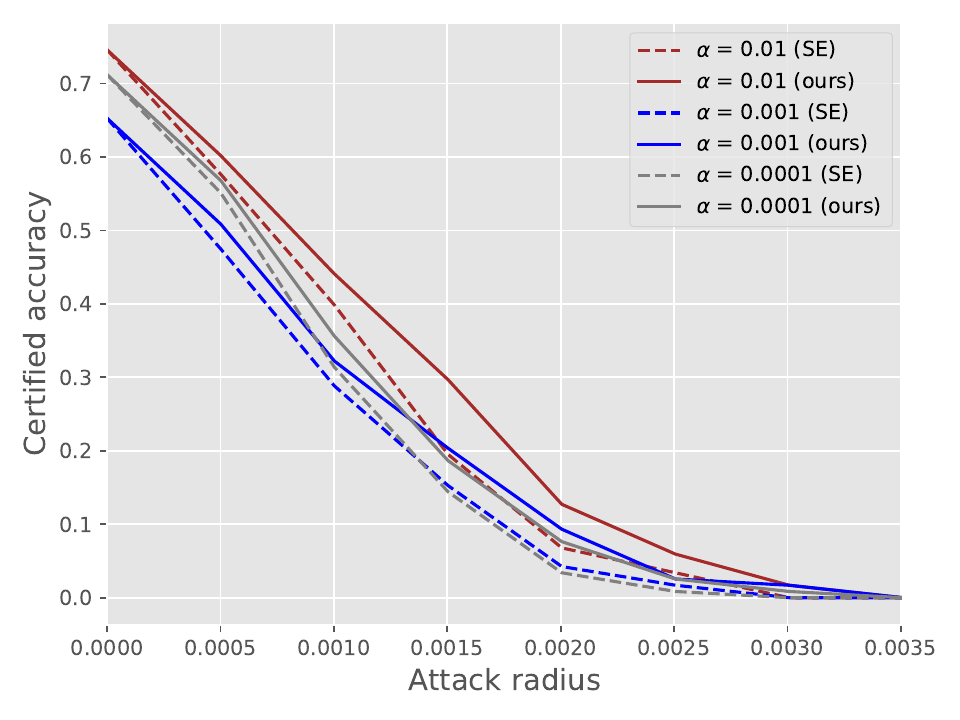}
         \caption{Dependency on $\alpha$.}
     \end{subfigure}
     \begin{subfigure}{0.31\textwidth}
         \centering
         \includegraphics[width=\textwidth]{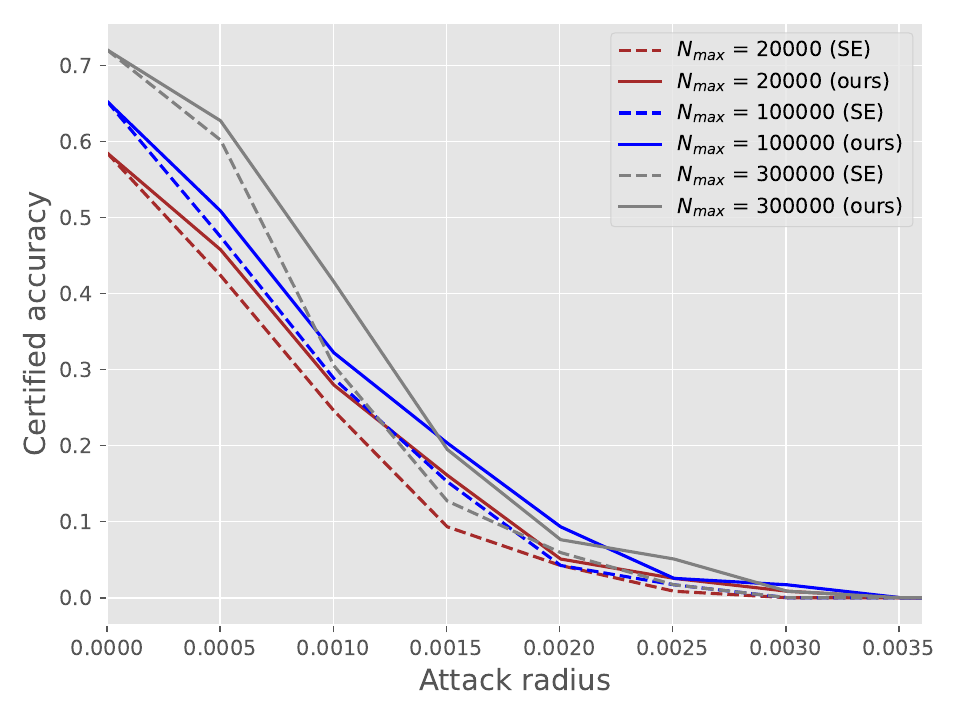}
         \caption{Dependency on $N_{\text{max}}$.}
     \end{subfigure}
        \caption{Pyannote model. Few-shot setting. Dependency of certified accuracy on the variance $\sigma$ of the additive noise, confidence level $\alpha$, and maximum number of noise samples $N_{\text{max}}$.} 
        \label{fig:pyannote_fewshot_1}
\end{figure*}
\begin{figure*}[htb]
     \centering
     \begin{subfigure}{0.31\textwidth}
         \centering
         \includegraphics[width=\textwidth]{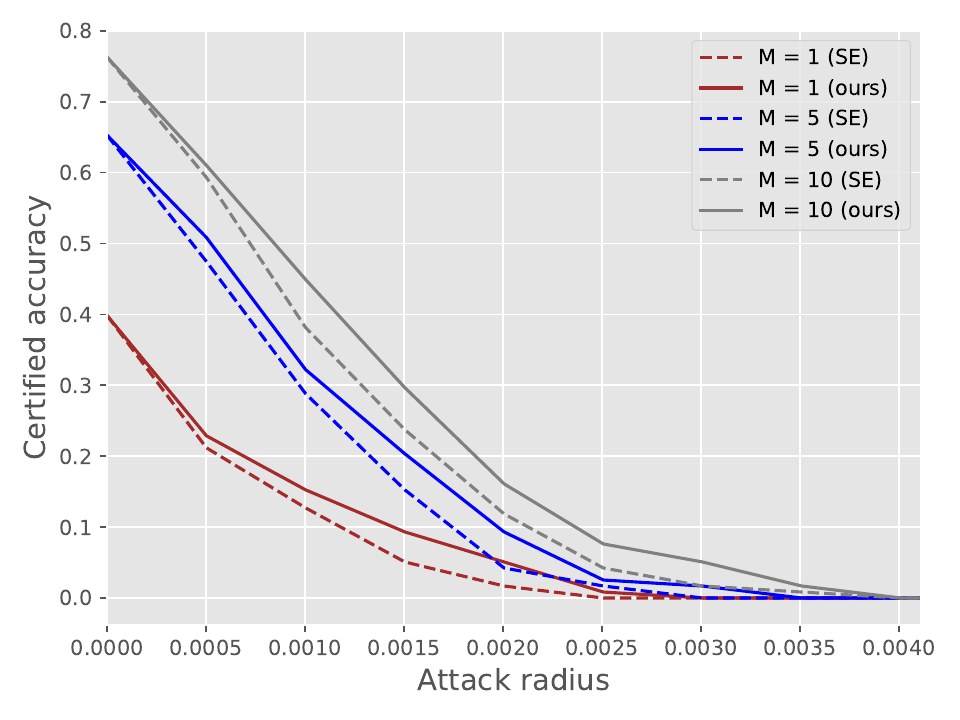}
         \caption{Dependency on $M$.}
     \end{subfigure}
     \begin{subfigure}{0.31\textwidth}
         \centering
         \includegraphics[width=\textwidth]{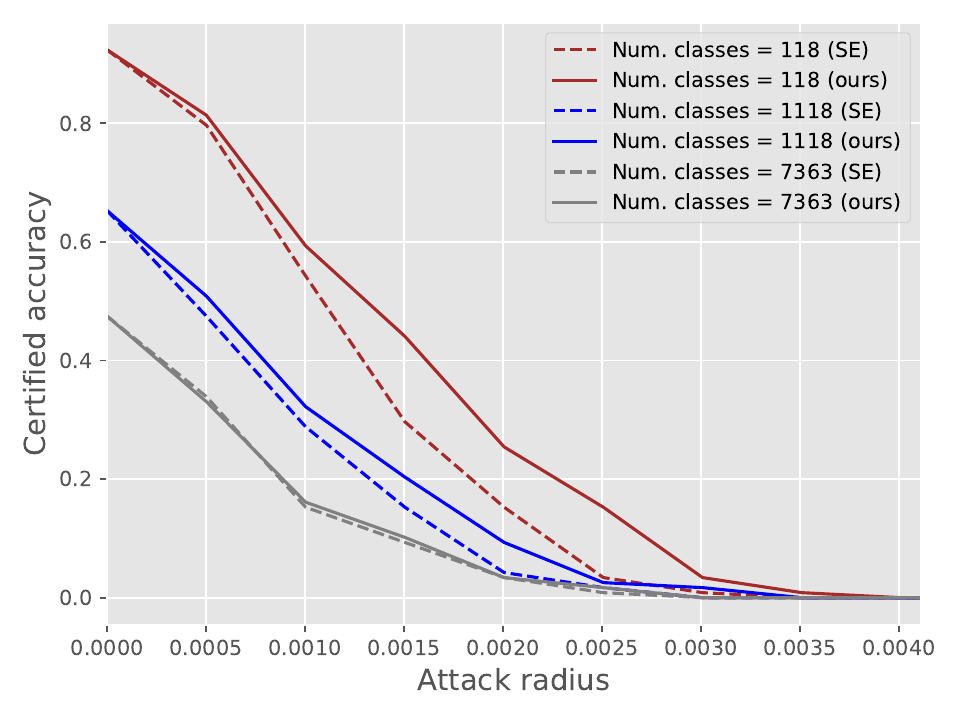}
         \caption{Dependency on number of speakers.}
     \end{subfigure}
     \begin{subfigure}{0.31\textwidth}
         \centering
         \includegraphics[width=\textwidth]{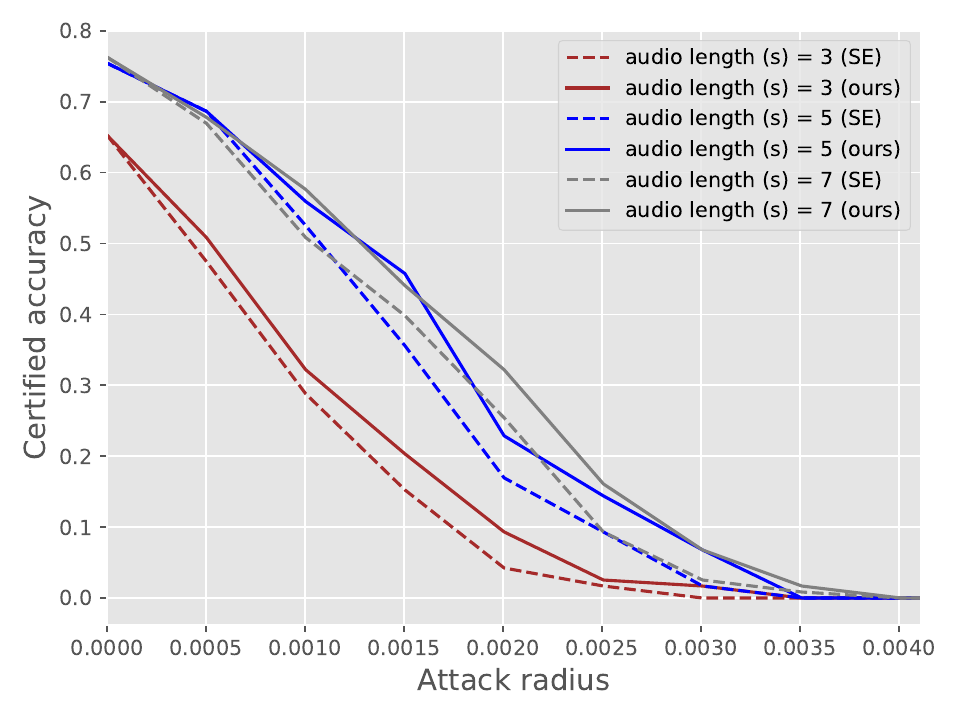}
         \caption{Dependency on the audio length.}
         \label{fig:pyannote_on_time}
     \end{subfigure}
        \caption{Pyannote model. Few-shot setting. Dependency of certified accuracy on number  $M$ of audios of a single speaker, number of enrolled speakers $K$, and the audio length in seconds.}
        \label{fig:pyannote_fewshot_2}
\end{figure*}

In this section, we describe the numerical implementation of the proposed method. 

\subsection{Sample Mean Instead of Expectation}
Notably, the prediction of the smoothed model from Eq. \eqref{eq:rs_model_smooth} is an expected value of the random variable that is the function of the base classifier. Hence, it is impossible to evaluate it exactly in the case of nontrivial $f$. Consequently, evaluating the mapping $\phi$ from Eq. \eqref{eq:phi} is impossible. A conventional way to deal with this problem is to replace the smoothed model with its unbiased estimation -- sample mean computed over $N$ samples, namely
\begin{equation}
    \hat{g}(x) = \frac{1}{N} \sum_{i=1}^N f(x+\varepsilon_i),
    \label{eq:g_mc}
\end{equation}
where $\varepsilon_i$ are the independent identically distributed normal random variable. However, it is also impossible to exactly determine which two centroids $c_{i_1}$ and $c_{i_2}$ are the closest ones to the true value of the smoothed classifier from Eq.~\eqref{eq:rs_model_smooth}.  We solve the issues mentioned above in the following manner:
\begin{enumerate}
    \item Firstly, we compute interval estimations of the distances between $g(x)$ and all the centroids using Hoeffding inequality \cite{hoeffding1994probability}. It is done to determine the two closest centroids with sufficient confidence.
    \item Secondly, given the two closest centroids, we compute the lower confidence bound $\hat{\phi}$ of $\phi$ from Eq.~\eqref{eq:phi}.
    \item Finally, when $\hat{\phi}$ is computed, the value $R(\hat{\phi}, \sigma)$ from Theorem \ref{th:robustness_guarantee} is treated as  the certified radius of $g$ at $x.$
\end{enumerate}

\begin{algorithm}[t]
    \textbf{Input}: $f, x$\\
    \textbf{Parameter}: $N, N_{\text{max}}, \sigma, \alpha$\\
    \textbf{Output}: R
    \begin{algorithmic}[1]
        \State{$\operatorname{isFinished} \gets \operatorname{False}$}
        \State{$N_0 \gets N$}
        \While{$\operatorname{not}\operatorname{isFinished}$ or $N \le N_{\text{max}}$}
        \State{$\varepsilon_1, \dots, \varepsilon_N \sim \mathcal{N}(0, \sigma^2 I)$}
        \State{$\varepsilon_{N+1}, \dots, \varepsilon_{2N} \sim \mathcal{N}(0, \sigma^2 I)$}
        \State{$\hat{g}_1(x) = \frac{1}{N}\sum_{j=1}^N f(x+\varepsilon_j)$}
        \State{$\hat{g}_2(x) = \frac{1}{N}\sum_{j=1}^N f(x+\varepsilon_{N+j})$}
        \For{$i \in \{1, \dots, K\}$}
        \State{$v^1_i = \hat{g}_1(x) - c_i$}
        \State{$v^2_i = \hat{g}_2(x) - c_i$}
        \State{$(l_i, u_i) \gets \textsc{HoeffdingCI}(v^1_i, v^2_i, \alpha)$}
        \Comment{Computation of two-sided CI using Hoeffding inequality, namely $(l_i, u_i): \mathbb{P}\left( \|{g}(x) - c_i\|_2  \in (l_i, u_i)\right) \ge 1 - \alpha$}
        \EndFor{}
        \State{$i_1 \gets \argmin\{l_1,\dots, l_K\}$}
        \State{$i_2 \gets \argmin\{l_1, \dots, l_K \setminus l_{i_1}\}$}
        \State{$i_q \gets \argmin\{l_1, \dots, l_K \setminus \{l_{i_1}, l_{i_2}\}\}$}
        \If{$u_{i_1} < l_{i_2} \wedge u_{i_2} < l_{i_q}$}
        \State{$\operatorname{isFinished} \gets \operatorname{True}$}
        \State{$\hat{g}(x) = \frac{\hat{g}_1(x) + \hat{g}_2(x)}{2}$}
        \State{$\tilde{\phi} \gets \frac{\langle \hat{g}(x), c_{i_1}-c_{i_2} \rangle}{2\|c_{i_1}-c_{i_2}\|_2} + \frac{1}{2}$}
        \State{$\hat{\phi} \gets \textsc{HoeffdingLowerBound}(\tilde{\phi}, \alpha)$}
        \State{$R \gets \sigma \Phi^{-1}(\hat{\phi})$}
        \State{\Return R}
        \Else
        \If{$2N > N_{\text{max}}$}
        \State{\Return Abstain}
        \Else
        \State{$N \gets N + N_0$}
        \EndIf
        \EndIf
        \EndWhile{}
    \end{algorithmic}
    \caption{Computation of the certified radius.}
    \label{alg:cr}
\end{algorithm}

\subsection{Hoeffding Confidence Interval and Error Probability}

Hoeffding inequality \cite{hoeffding1994probability} bounds the probability of a large deviation of a sample mean from the population mean, namely
\begin{equation}\label{eq:hoeffding}
    \mathbb{P}(|\overline{X} - \mathbb{E}(X)| \ge t) \le 2 \exp \left(-\frac{2t^2 N^2}{\sum_{i=1}^N (b_i-a_i)^2} \right),
\end{equation}
where $\overline{X} = \frac{1}{N}\sum_{i=1}^N X_i,$ and $X_i$ are i.i.d. random variables such that $\mathbb{P}(X_i \in (a_i, b_i)=1).$


\subsubsection{Distances to the Centroids.}
An estimation of distance between the smoothed embedding $g(x)$ from Eq. \eqref{eq:rs_model_smooth} and the speaker centroid $c_i$ from Eq. \eqref{eq:class_prot} may be derived from an estimation of the dot product $\langle \hat{g}_1(x) -c_i, \hat{g}_2(x) - c_i \rangle,$ where
\begin{equation}\label{eq:two_estimates}
\begin{aligned}
    \hat{g}_1(x) &= \frac{1}{N}\sum_{i=1}^N f(x+\varepsilon_i), \\
    \hat{g}_2(x) &= \frac{1}{N}\sum_{j=1}^N f(x+\varepsilon_j)
\end{aligned}
\end{equation}
are two independent unbiased estimates of $g(x).$  
Once computed, confidence interval $(l^2_i, u^2_i)$ for the expression $\langle \hat{g}_1(x) -c_i, \hat{g}_2(x) - c_i \rangle$ implies confidence interval $(l_i, u_i)$ of interest. The work of \cite{pautov2022smoothed} provides a detailed derivation of the confidence interval.

\subsubsection{Estimation of $\hat \phi$.} Hoeffding inequality is also used to compute confidence intervals for the value $\phi$ from Theorem \ref{th:robustness_guarantee}. 
Namely, given 
\begin{equation}\label{eq:phi_hat}
    \tilde{\phi} -\frac{1}{2}= \frac{\left\langle \frac{\hat{g}_1(x) + \hat{g}_2(x)}{2}, c_{i_1} - c_{i_2} \right\rangle}{2\|c_{i_1}-c_{i_2}\|_2},
\end{equation}
as an estimation of $\phi - \frac{1}{2}$ over $2N$ samples $\xi_j$ in the form
\begin{equation}
   \xi_j = \frac{\langle f(x+\varepsilon_j), c_{i_1}-c_{i_2}\rangle}{2\|c_{i_1}-c_{i_2}\|_2},
\end{equation}
such that $\xi_j \in [-\frac{1}{2}, \frac{1}{2}],$ we compute lower bound $\hat{\phi}-\frac{1}{2}$ of $\phi - \frac{1}{2}$ in the form
\begin{equation}\label{eq:phi_bound}
    \hat{\phi}-\frac{1}{2} = \tilde{\phi} - \frac{1}{2} - \sqrt{\frac{\ln{\frac{2}{\alpha}}}{4N}}.
\end{equation}
Note that $\alpha$ in Eq. \eqref{eq:phi_bound} is the upper bound for the error probability. In other words, 

\begin{equation}
    \mathbb{P}(\phi < \hat{\phi}) < \alpha. 
\end{equation}
All the procedures are combined in the numerical pipeline presented in the Algorithm \ref{alg:cr} and schematically in Fig.~\ref{fig:main_fig}.

\subsubsection{Error Probability of  Algorithm \ref{alg:cr}.} 

Since the procedure in Algorithm \ref{alg:cr} is not deterministic (as it depends on the computation of confidence intervals), it is important to estimate its failure probability. First, estimating the two closest centroids is statistically sound only if all the distances between smoothed embedding and the centroids are within corresponding confidence intervals. In contrast, if at least one of the distances
\begin{equation}
    \|g(x)-c_1\|_2, \dots, \|g(x)-c_K\|_2
\end{equation}
is not within the corresponding interval, it is impossible to guarantee that the two closest centroids are correctly determined. Thus, all the respective Hoeffding inequalities have to hold. It happens with the probability $p_1 = (1-\alpha)^K,$ where $K$ is the number of classes. Secondly, note that the lower confidence bound for $\phi$ from Theorem \ref{th:robustness_guarantee} is correct with probability $p_2 = (1-\alpha).$ 

Thereby, the probability of the correct output of Algorithm \ref{alg:cr} is $p_1 p_2 = (1-\alpha)^{K+1}$ what leads to the error probability $q = 1 - (1-\alpha)^{K+1}.$

\section{Experiments}

\subsection{Datasets}
For our experiments, we used the VoxCeleb1 \cite{nagrani2017voxceleb} and VoxCeleb2 \cite{chung2018voxceleb2} datasets, which are standard for speaker recognition and verification tasks. VoxCeleb1 comprises $1211$ development speakers and $40$ test speakers, with over $150000$ utterances spanning $350$ hours. VoxCeleb2 includes $5994$ development speakers and $118$ test speakers, totaling about $2400$ hours and $1.1$ million utterances. These multilingual datasets feature speakers from over $140$ nationalities, covering various accents and ages. We evaluated our method by varying the number of enrolled speakers from $118$ (VoxCeleb2 test set) to nearly all available speakers ($7323$), excluding the VoxCeleb1 test set.

\subsection{Evaluation Protocol}

\begin{figure*}[htb]
     \centering
     \begin{subfigure}{0.31\textwidth}
         \centering
         \includegraphics[width=\textwidth]{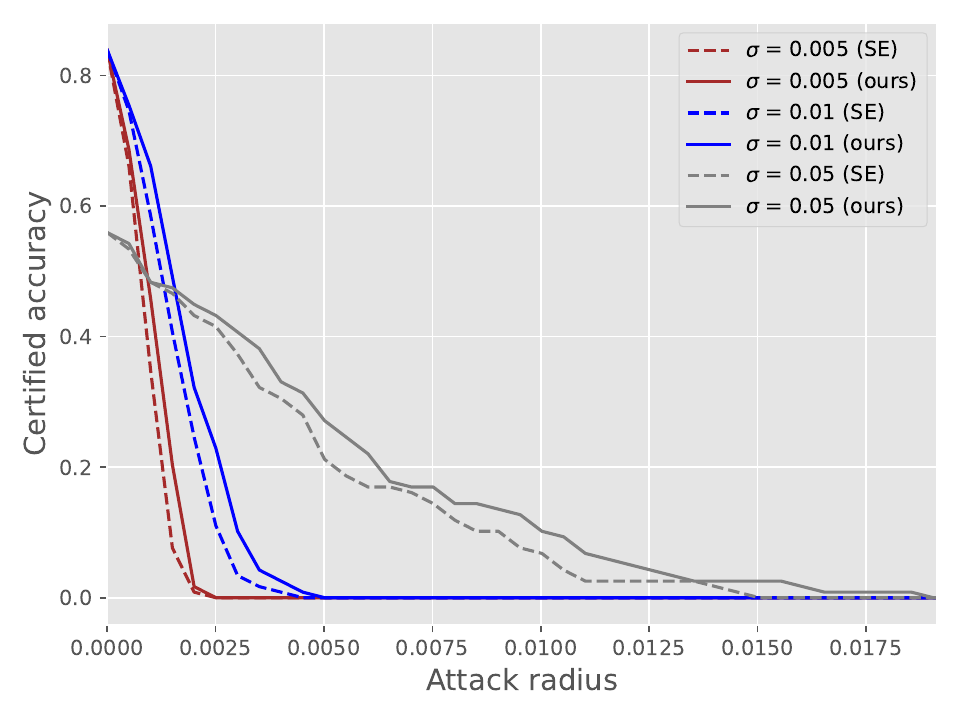}
         \caption{Dependency on $\sigma$.}
     \end{subfigure}
     \begin{subfigure}{0.31\textwidth}
         \centering
         \includegraphics[width=\textwidth]{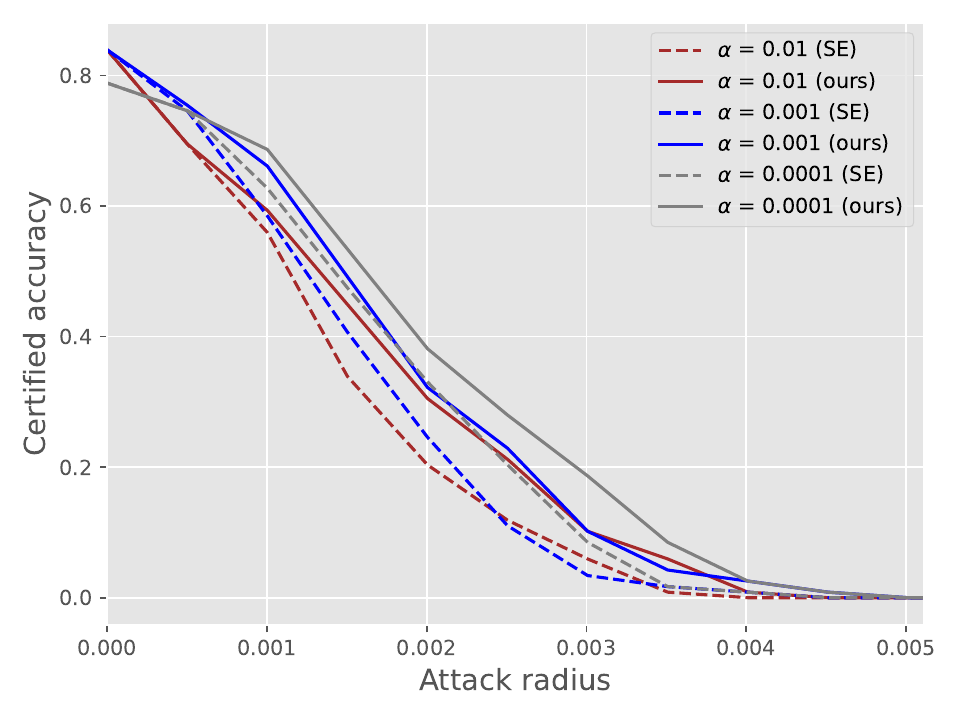}
         \caption{Dependency on $\alpha$.}
     \end{subfigure}
     \begin{subfigure}{0.31\textwidth}
         \centering
         \includegraphics[width=\textwidth]{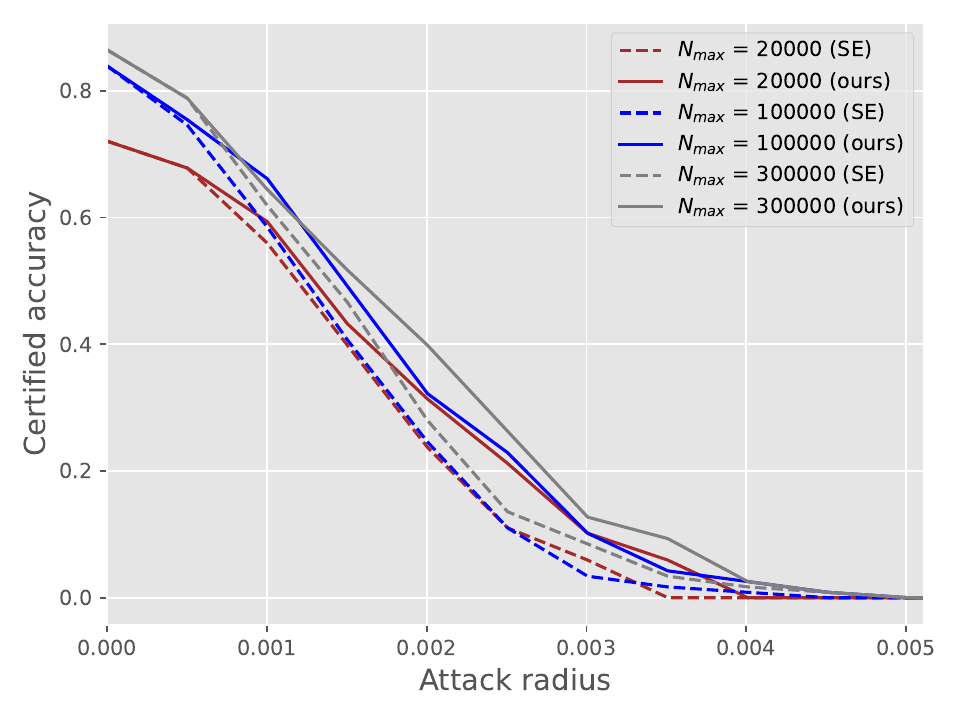}
         \caption{Dependency on $N_{\text{max}}$.}
     \end{subfigure}
        \caption{ECAPA-TDNN model. Few-shot setting. Dependency of certified accuracy on the variance $\sigma$ of the additive noise, confidence level $\alpha$, and maximum number of noise samples $N_{\text{max}}$.} 
        \label{fig:ecapa_fewshot_1}
\end{figure*}
\begin{figure*}[htb]
     \centering
     \begin{subfigure}{0.31\textwidth}
         \centering
         \includegraphics[width=\textwidth]{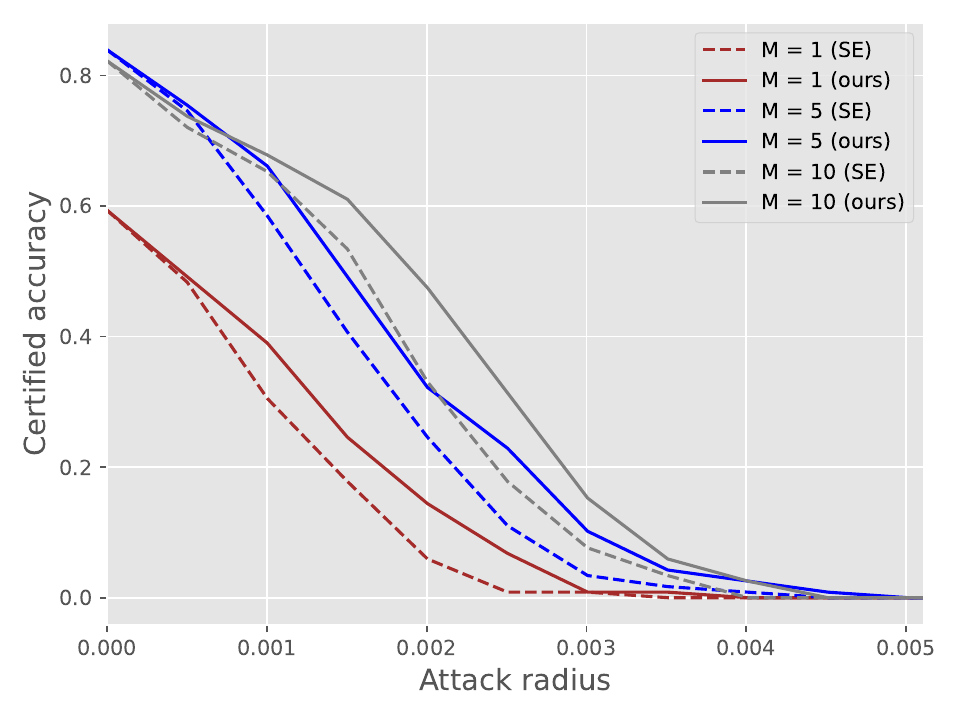}
         \caption{Dependency on $M$.}
     \end{subfigure}
     \begin{subfigure}{0.31\textwidth}
         \centering
         \includegraphics[width=\textwidth]{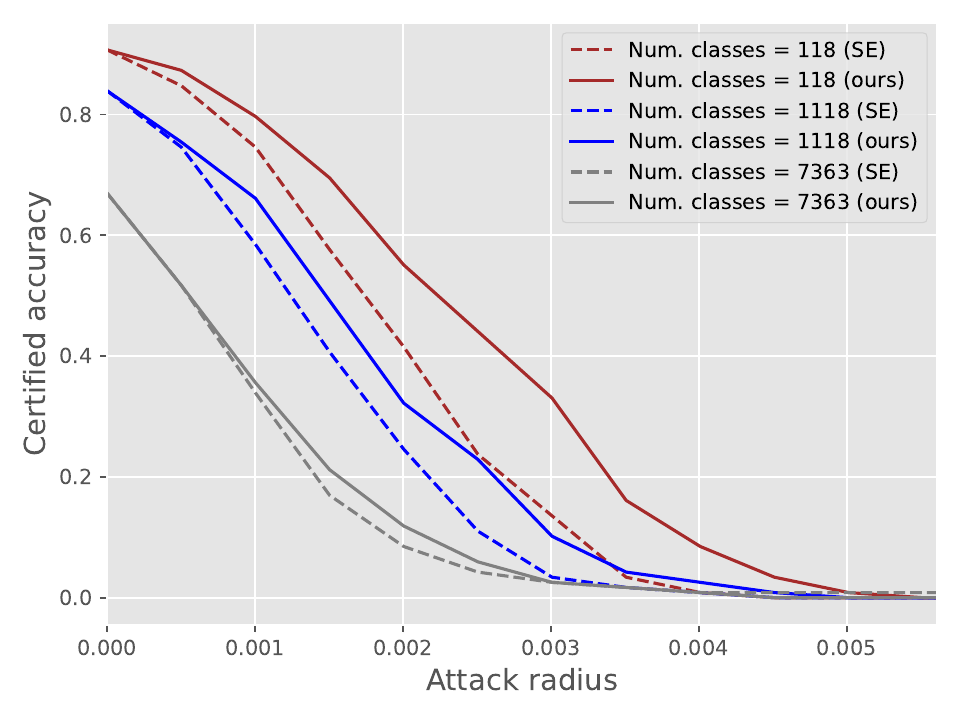}
         \caption{Dependency on number of speakers.}
     \end{subfigure}
     \begin{subfigure}{0.31\textwidth}
         \centering
         \includegraphics[width=\textwidth]{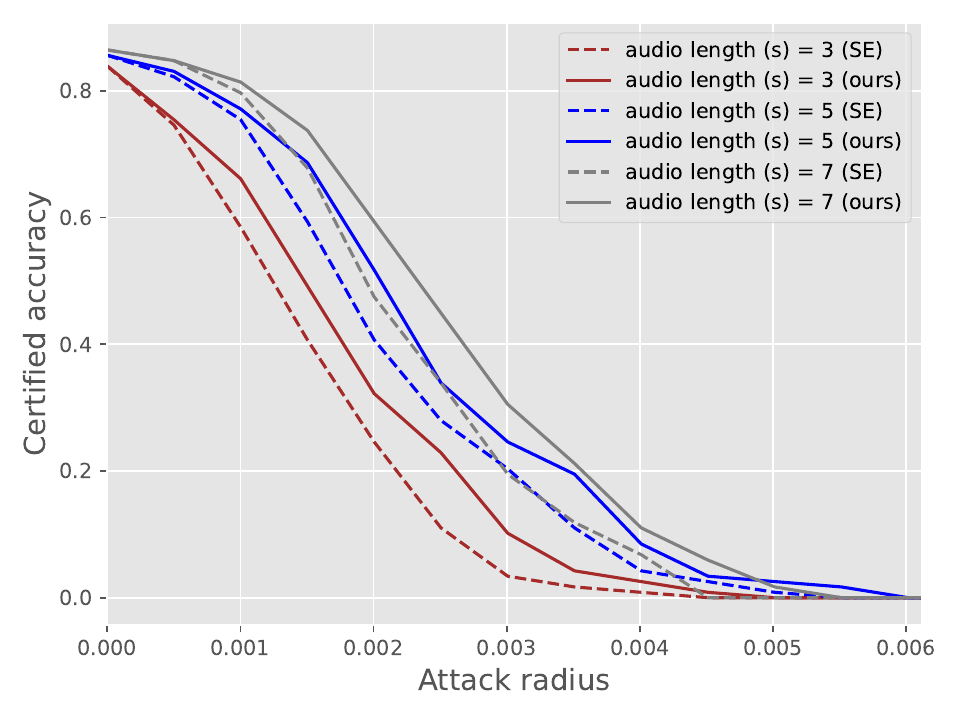}
         \caption{Dependency on the audio length.}
         \label{fig:ecapa_on_time}
     \end{subfigure}
        \caption{ECAPA-TDNN model. Few-shot setting. Dependency of certified accuracy on number  $M$ of audios of a single speaker, number of enrolled speakers $K$, and the audio length in seconds.}
        \label{fig:ecapa_fewshot_2}
\end{figure*}

\begin{figure*}[!ht]
     \centering
     \begin{subfigure}{0.31\textwidth}
         \centering
         \includegraphics[width=\textwidth]{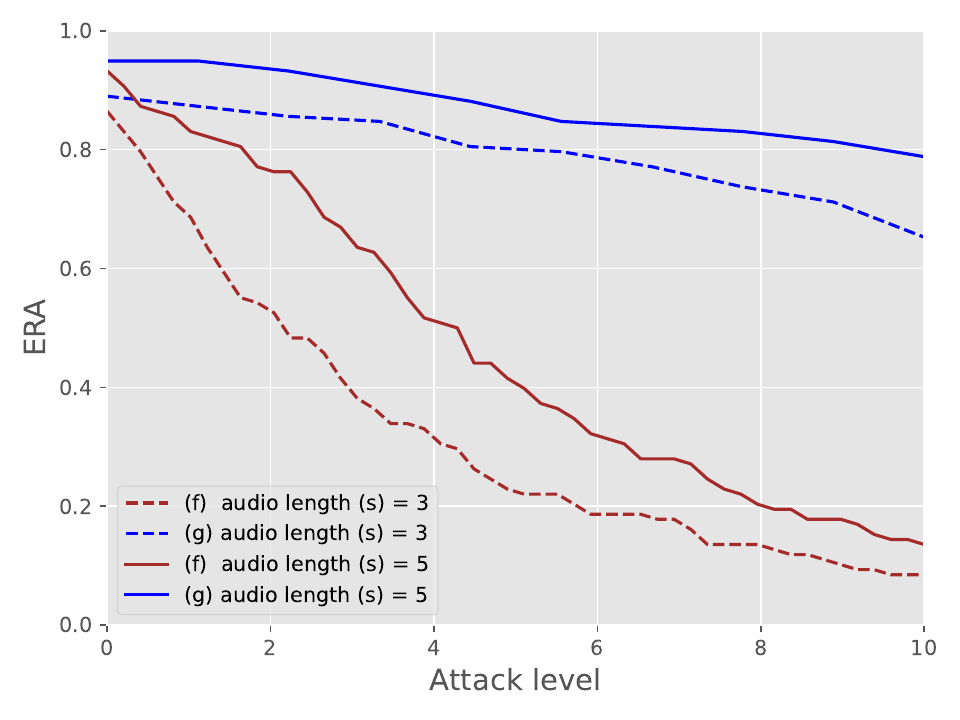}
         \caption{Gaussian noise.}
         \label{fig:era_gaus}
     \end{subfigure}
     \begin{subfigure}{0.31\textwidth}
         \centering
         \includegraphics[width=\textwidth]{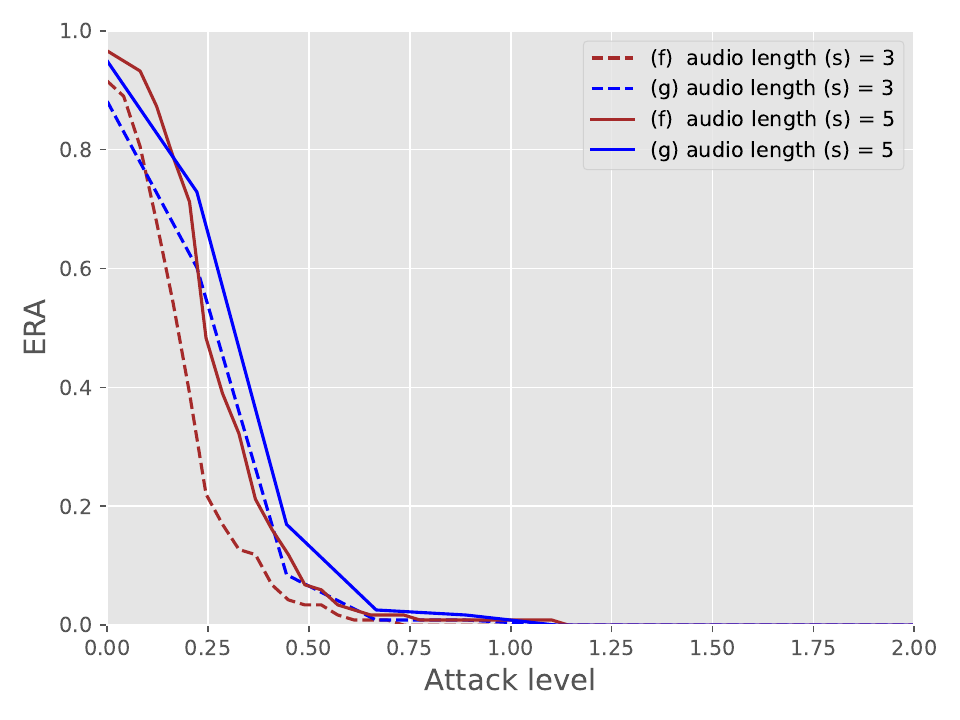}
         \caption{PGD.}
         \label{fig:era_pgd}
     \end{subfigure}
     \begin{subfigure}{0.31\textwidth}
         \centering
         \includegraphics[width=\textwidth]{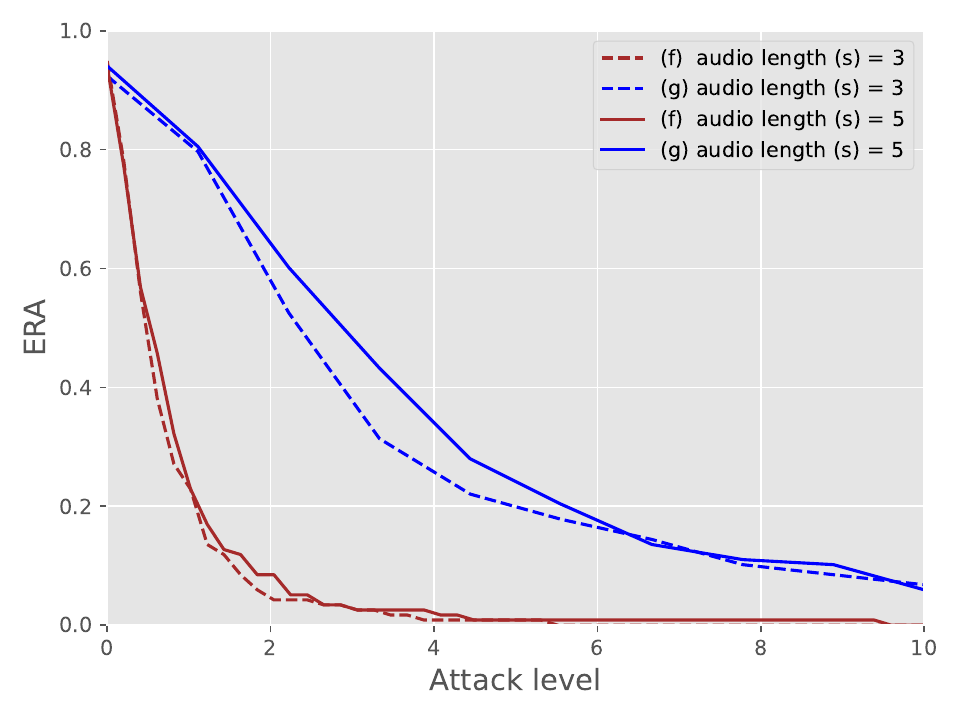}
         \caption{UAP.}
         \label{fig:era_anon}
     \end{subfigure}
    \caption{Pyannote model. Few-shot setting. Empirically Robust Accuracy of $f$ and $g$ in the presence of additive perturbations: Gaussian noise, PGD adversarial attack, speaker anonymization Universal Adversarial Patch (UAP) \cite{liu2024transferable}.}
        \label{fig:era_time}
\end{figure*}

We evaluate the methods in several settings. Experiments were conducted using various backbone embedding models: ECAPA-TDNN \cite{desplanques2020ecapa} from the Speechbrain framework \cite{ravanelli2021speechbrain} that utilizes Mel-Spectrogram for the frontend; the Pyannote framework \cite{bredin2020pyannote}, which focuses on speaker diarization and utilizes the raw-waveform frontend SincNet. These models transform speech into vector representations of dimensions $d =$ 192 and 512 correspondingly. For the ECAPA-TDNN-based $f$, plain accuracy is $\operatorname{Acc} = 95.0\%$, equal-error-rate $\operatorname{EER}(f) = 0.34\%$, $\operatorname{EER}(g) = 0.89\%$. For the Pyannote-based $f$, $\operatorname{Acc} = 88.3\%$, $\operatorname{EER}(f) = 1.17\%$, $\operatorname{EER}(g) = 1.40\%$. $\operatorname{EER}$ is a decision threshold regarding the ASV or classification task for which the model's false acceptance and false rejection rates are equal. We conducted experiments in an ASI setting.


For the certification procedure in Algorithm \ref{alg:cr}, the default parameters are the following: standard deviation of additive noise used for smoothing $\sigma = 10^{-2}$, the maximum number of samples to construct $\hat{g}$ is set to be $N_{\text{max}} = 10^5$, the confidence level $\alpha=10^{-3}$, number of enrolled speakers is $K=1118$, number of random audios used to create the speaker enrollment vector $M=5$, and length of given audios is set to be 3s with sampling rate 16 kHz, number of speakers in the test set $S^i$ is 118 (VoxCeleb2 Test).

For the evaluation, we considered $K$ enrolled speakers and, for each of them, created $c_k \in S^c$ of $M$ randomly sampled speaker's enrollment audios, which are presented in $S^e$. We tested our models, providing inference audios $x \in S^i, ~ S^e \cap S^i = \emptyset,$ where number of unique test speakers in $S^i$ is fixed and equal to 118 (VoxCele2 test). We report certified accuracy (CA) for each method on the $S^c$ centroids and $S^i$ test audios. Certified accuracy represents the proportion of correctly matched samples from $S^i$ to the corresponding centroids in $S^c$ for which the smoothed model has a certified radius exceeding the given attack magnitude. Specifically, given the recognition rule
\begin{equation}\label{eq:class_rule}
    i_1(x) = \argmin_{k \in \{1, \dots, K\}} \rho \left(g(x), c_k\right),
\end{equation}
and the norm of perturbation $\varepsilon,$ the certified accuracy is computed as follows:
\begin{equation}\label{eq:cert_accuracy}
    CA(S^c, S^i, \varepsilon) = \frac{|(x,y) \in S^i: R(x) > \varepsilon\ \wedge \ i_1(x) =y|}{|S^i|},
\end{equation}
where $R(x)$ is the certified radius from Theorem~\ref{th:robustness_guarantee}.

We compared our approach to the work of \cite{pautov2022smoothed}, where authors propose the method called \textit{Smoothed Embeddings} (\textit{SE}) to certify prototypical networks. The certified radius $R^{SE}(x)$ produced by \textit{SE} has the form 
\begin{equation}\label{eq:ca_se}
    R^{SE}(x) = \sqrt{\frac{\pi \sigma^2}{2}} \frac{\|c_{i_2} - g(x)\|_2^2 - \|c_{i_1}-g(x)\|_2^2}{2\|c_{i_1}-c_{i_2}\|^2_2}. 
\end{equation}
in our notation. In contrast to our work, they perform a geometrical analysis of Lipschitz properties of the smoothed model,  whereas we study the properties of the scalar mapping from the embedding space. 

We also provided results (see Appendix of the full version of the manuscript) based on vanilla RS \cite{cohen2019certified, salman2019provably} in the classification setting~\eqref{eq:cert_r} of a fixed number $K$ of speakers in $S^c$. Default parameters were the same as in our approach. Since an exact estimation of $g_{\text{clf}}(x)$ \eqref{eq:rs_model} is impossible, a similar sample-mean is utilized with the Clopper-Pearson \cite{clopper1934use} test for estimation of $\hat p_{i_1}$ which is the lower confidence bound of $p_{i_1}$. In a nutshell, this is a Binomial proportion test confidence interval of top class v.s. the rest. This requires the correct class to be predicted in more than half of samples $\hat p_{i_1} > \frac{1}{2}$ for the certification.


The computational time is approximately $30$ seconds for the Pyannote model and $120$ seconds for ECAPA-TDNN.

\subsection{Results and Discussion}

In Figures \ref{fig:pyannote_fewshot_1}, \ref{fig:pyannote_fewshot_2} and \ref{fig:ecapa_fewshot_1}, \ref{fig:ecapa_fewshot_2} we present results that illustrate the effects of varying a single parameter while keeping all other at their default values for the SE and our approaches for two backbone models. Several observations can be obtained from these results:
\begin{itemize}
    \item $\sigma$ significantly impacts the certification system (ours, SE, and RS). Higher values lead to a more robust system, which comes at the expense of reduced accuracy (robustness-accuracy trade-off);
    \item $\alpha$ does not affect the certification significantly; 
    \item There are threshold values for the number of speaker enrollment audios $M$ and audio length beyond which the results remain nearly unchanged;
    \item Evidently, an increase of $N_{\text{max}}$ parameter enhances the certification process, while classification difficulty rises as the number of enrolled speakers $K$ increases.
\end{itemize}

Additionally, our method demonstrates a marginal improvement across all scenarios compared to the SE approach. Figures \ref{fig:pyannote_fewshot_1} - \ref{fig:ecapa_fewshot_2} illustrate that our method achieves enhanced certified accuracy for the same attack levels. 

In Figure \ref{fig:era_time}, we demonstrate empirical robust accuracy (ERA) -- the fraction of correctly recognized perturbed audios $x + \delta$ for all sampled perturbations $\delta \leq l,$ where $l$ is a current attack level. $g$ was estimated as in Eq. \eqref{eq:g_mc} without certification criteria. Projected Gradient Descent \cite{madry2017towards} is selected as it is considered a standard adversarial attack to evaluate models' robustness. One can notice that the empirical robustness of $g$ and $f$ is significantly better than the certification results of $g$.  Nonetheless, presented attacks do not necessarily convey the worst certification result, as stronger attacks exist, and the worst-case ERA might be closer to CA.


The certification condition in Theorem~\ref{th:robustness_guarantee} does not depend on audio length explicitly. For a given sample $x \in \mathbb{R}^n$, it yields the certified radius $R(x)$ -- a lower bound on the $l_2-$norm of a perturbation $\delta$ that can change a smoothed model's prediction. However,  the relative distortion (e.g., signal-to-noise ratio) differs for various $n$. One can notice from the  (Fig.~\ref{fig:pyannote_on_time} and Fig.~\ref{fig:ecapa_on_time}) that the longer the audio sample is, the smaller the relative distortion is and consequently, certification results are better. Additionally, achieving audio-length independent certification against $l_{\infty}-$norm bounded perturbations seems unsolvable \cite{hayes2020extensions}. The Theorem~\ref{th:robustness_guarantee} is still valid if $l_2-$norm as the distance function is replaced by the negative cosine distance.

Our method is evaluated for the speaker identification task only. The method can be transferred to the speaker diarization task but cannot be applied directly in an ASV scenario.

It is feasible to extend our certification procedure to multiplicative and semantic transformations \cite{muravevcertified, li2021tss} by applying different mappings and smoothing distributions. Nonetheless, the method certifies the model only against additive perturbations for the fixed voiceprint $x$, but these guarantees do not apply a priori to the new voiceprint $x_1$ even if it is a genuine speech of the same speaker: $\forall \delta: \|\delta\|_2 \leq R(x) \mapsto \argmin_k \rho(g(x + \delta), c_k)=i_1$, where $i_1$ is a correct class, but $\exists \delta_1: \| \delta_1 \|_2 \leq R(x),$ but $\argmin_k \rho(g(x_1 + \delta_1), c_k) \ne i_1$. Additionally, current methods cannot help to certify SR models against rapidly evolving deepfakes \cite{yamagishi2021asvspoof, wang2024asvspoof5}. 

Although RS over class probabilities provides better certification radii (see Appendix of the full version of the manuscript) compared to our approach, our method does not imply knowledge of the class probabilities that may be more suitable for the metric learning tasks. 

\section{Conclusion}
In this work, we presented a new approach to certify speaker identification models that map input audios to normalized embeddings against norm-bounded additive perturbations. We introduced scalar mapping from the embedding space and derived theoretical robustness guarantees based on its Lipschitz properties. We experimentally evaluated our approach against the concurrent method and achieved state-of-the-art certification results in a few-shot setting. In addition, our method can be applied to the certification of other metric learning tasks, such as face biometrics. 

In summary, we expect this work to highlight the issue of certified robustness in biometrics systems, particularly in speaker identification, and improve AI safety. Future developments in this topic might be devoted to improving empirical and certified guarantees and developing certification against other types of attacks, including non-additive ones such as deepfakes.

\section*{Acknowledgements}
The authors acknowledge the support from the Russian Science Foundation grant No. 25-41-00091. The authors are grateful to Olesya Kuznetsova for valuable discussions during the preparation of this paper.


\bibliography{aaai25}
\appendix
\onecolumn
\section{Appendix}

\subsection{Proof of the Theorem}
In this section, we provide the proof of  Theorem \ref{th:robustness_guarantee}. 

\setcounter{theorem}{0}
\begin{theorem}[Restated]\label{th:robustness_guarantee_restated}
Let $g$ be the model from Eq. \eqref{eq:rs_model_smooth} and $c_1, \dots, c_K$ be the class prototypes from Eq. \eqref{eq:class_prot}. Suppose that audio $x$ is correctly assigned to class $c$ represented by prototype $c_{i_1}$ and $c_{i_2}$ is the second closest to $g(x)$ prototype. Then for all additive perturbations $\delta: \|\delta\|_2 \le R(\phi, \sigma) = \sigma  \Phi^{-1} (\phi),$
\begin{equation}\label{eq:rob_guarantee}
        \arg\min_{k \in [1,\dots K]} \|g(x) - c_k\|_2 =  \arg\min_{k \in [1,\dots K]} \|g(x+\delta) - c_k\|_2,
\end{equation}
where $\phi = \phi(x) = \frac{\langle g(x), c_{i_1}-c_{i_2}\rangle}{2 \|c_{i_1}-c_{i_2}\|_2} + \frac{1}{2}.$
\end{theorem}
\begin{proof}
For simplicity, let $\sigma=1.$ Consider the function $\psi(x) = \langle 2 g(x), c_{i_1}-c_{i_2}\rangle $ with the gradient 
\begin{align}\label{eq:proof_gr}
    &\nabla_x \psi(x) = \langle 2 \nabla_x g(x), c_{i_1}-c_{i_2}\rangle, \\
    &\nabla_x \psi(x) = 2 \langle \int_{\mathbb{R}^n} f(x+\varepsilon) \varepsilon \rho(\varepsilon) d\varepsilon, c_{i_1}-c_{i_2} \rangle = \int_{\mathbb{R}^n} r(\varepsilon) \varepsilon \rho(\varepsilon) d\varepsilon,
\end{align}
where $\rho(\varepsilon) = \frac{1}{(2\pi)^{n/2}}\exp\left(-\frac{\|\varepsilon\|^2_2}{2}\right)$ and $r(\varepsilon) = 2\langle f(x+\varepsilon), c_{i_1}-c_{i_2}\rangle$.
Note that $r(\varepsilon) \in \left[-2\|c_{i_1}-c_{i_2}\|_2, 2\|c_{i_1}-c_{i_2}\|_2\right]$ and $$\int_{\mathbb{R}^n} r(\varepsilon)\rho(\varepsilon)d\varepsilon = 2\langle g(x), c_{i_1}-c_{i_2} \rangle = \psi(x).$$
Let us introduce $\hat{r}(\varepsilon) = \frac{r(\varepsilon)}{4\|c_{i_1}-c_{i_2}\|_2} + \frac{1}{2}$.
Note that $\hat{r}(\varepsilon) \in [0,1]$ and $\int_{\mathbb{R}^n} \hat{r}(\varepsilon)\rho(\varepsilon)d\varepsilon = \phi(x).$
The expression of gradient from Eq. \eqref{eq:proof_gr} takes the form 
\begin{equation}\label{eq:new_gr}
    \nabla_x \psi(x) = \int_{\mathbb{R}^n} \big[ 4\|c_{i_1}-c_{i_2}\|_2 \hat{r}(\varepsilon) - 2\|c_{i_1}-c_{i_2}\|\big] \varepsilon \rho(\varepsilon)d\varepsilon = 4\|c_{i_1}-c_{i_2}\|_2 \int_{\mathbb{R}^n}\hat{r}(\varepsilon)\varepsilon \rho(\varepsilon)d\varepsilon.
\end{equation}
To compute $\sup_{x} \left\|\nabla_x \psi(x)\right\|_2,$ we need to find \begin{align*}
    &\sup_{v: \|v\|_2=1} \mathbb{E}_{\varepsilon \sim \mathcal{N}(0, I)} \langle \hat{r}(\varepsilon)\varepsilon, v\rangle\\
    & \text{subject to}\ \mathbb{E}_{\varepsilon \sim \mathcal{N}(0, I)} \hat{r}(\varepsilon) = \phi(x).
\end{align*}
According to \cite{salman2019provably}, 

\begin{equation}
    \sup_{x} \left\|\nabla_x \psi(x)\right\|_2 = \sup_{v: \|v\|_2=1} \langle\nabla_x \psi(x), v\rangle = \frac{4\|c_{i_1}-c_{i_2}\|_2}{\sqrt{2\pi}}\exp \left[-\frac{1}{2} \left(\Phi^{-1}(\phi(x))\right)^2\right] = z(\psi(x)),
\end{equation}
where $\Phi^{-1}$ in the inverse of standard Gaussian CDF. Let's introduce the function $\xi = \xi(\psi(x)): \mathbb{R}^n \to \mathbb{R}^1$ such that 

\begin{equation}\label{eq:for_xi}
    \|\nabla_x \xi(\psi(x))\| \le \left|\frac{d \xi}{d \psi}\right|\left\|\nabla_x \psi(x)\right\| (x) \equiv 1
\end{equation}
and such that $\xi(\psi(x))$ is monotonically increasing, and, thus $\left|\frac{d\xi}{d\psi}\right| = \frac{d\xi}{d\psi}.$
\begin{equation}
\frac{d}{d x} \operatorname{erfc}^{-1}(x)=-\frac{1}{2} \sqrt{\pi} e^{\left[\operatorname{erfc}^{-1}(x)\right]^2}
\end{equation}
Note that the function 
\begin{equation}
    \xi(\psi(x)) = \int \frac{d \psi(x)}{z(\psi(x))} = \frac{\sqrt{2\pi}}{4\|c_{i_1}-c_{i_2}\|_2} \int \exp\left[\frac{1}{2} \left(\Phi^{-1}(\phi) \right)^2\right] d \psi(x) = \Phi^{-1}(\phi(x))
\end{equation}
satisfies Eq. \eqref{eq:for_xi}. \\
Thus, function $\xi(\psi(x)) = \Phi^{-1}(\phi(x))$ has Lipschitz constant $L \le 1,$ or, equivalently, $\forall \delta$

\begin{equation}\label{eq:for_r}
   \|\xi(\psi(x)) - \xi(\psi(x+\delta))\|_2 \le \|\delta\|_2.
\end{equation}
Since $x$ is correctly classified by $g$, $\psi(x)>0$ and $\phi(x) > \frac{1}{2}.$ Note that all the perturbations $\tilde{\delta}$ such that $\psi(x+\tilde{\delta}) = 0$ have norm not less than $\Phi^{-1}(\phi(x))$, since

\begin{align}
    \|\xi(\psi(x)) - \xi(\psi(x+\tilde{\delta}))\|_2 \le \|\tilde{\delta}\|_2 \Rightarrow \xi(\psi(x)) = \Phi^{-1}(\phi(x)) \le \|\tilde{\delta}\|. 
\end{align}
Consequently, all perturbations $\hat{\delta}$ such that $\psi(x+\hat{\delta}) \le 0$ have norm not less than $\Phi^{-1}(\phi(x))$ since $\psi(x)$ is continuous.\\

Finally, that means that for all perturbations $\delta: \|\delta\|_2 < \xi(\psi(x))=\Phi^{-1}(\phi(x)) \Rightarrow \psi(x)>0,$ what finalizes the proof. The proof for the cases when $\sigma \ne 1$ is analogous.

\end{proof}

\subsection{Additional Experiments}
In this section, we present the results of other experiments. Certified accuracy was calculated using the RS approach. The experimental settings are similar to those conducted for our method. 


\begin{figure*}[htb]
     \centering
     \begin{subfigure}{0.33\textwidth}
         \centering
         \includegraphics[width=\textwidth]{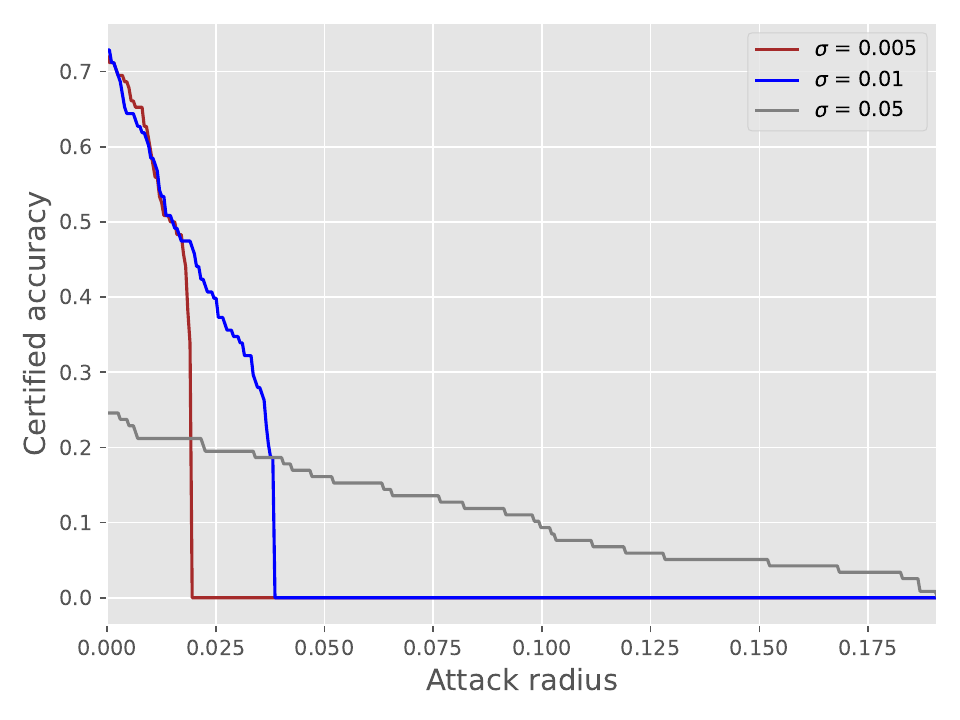}
         \caption{Dependency on $M$}
     \end{subfigure}
     \begin{subfigure}{0.33\textwidth}
         \centering
         \includegraphics[width=\textwidth]{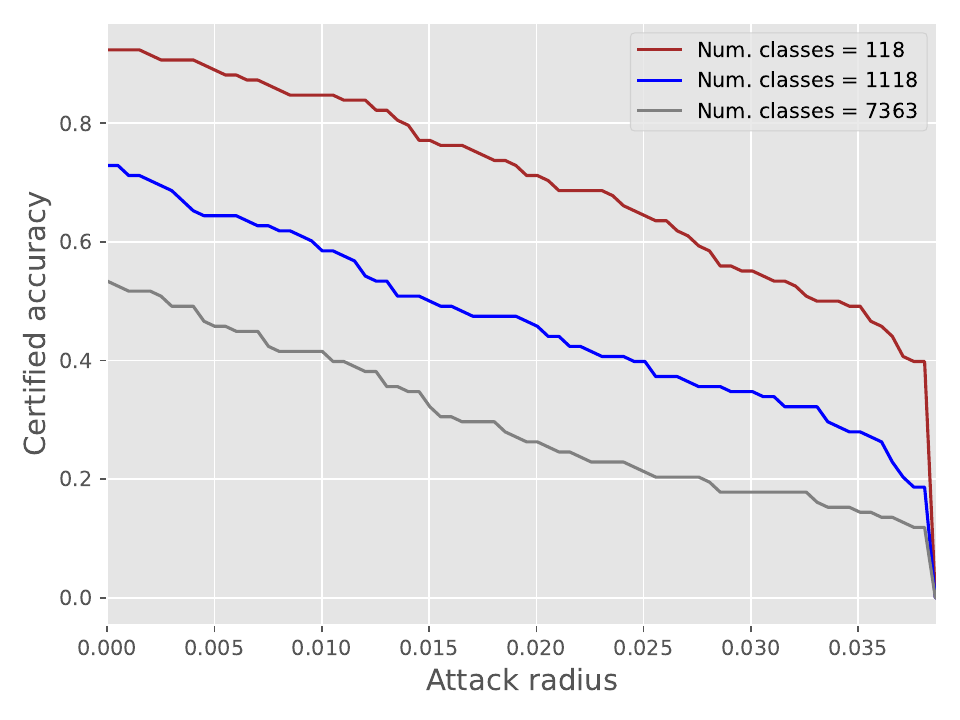}
         \caption{Dependency on number of speakers.}
     \end{subfigure}
     \begin{subfigure}{0.33\textwidth}
         \centering
         \includegraphics[width=\textwidth]{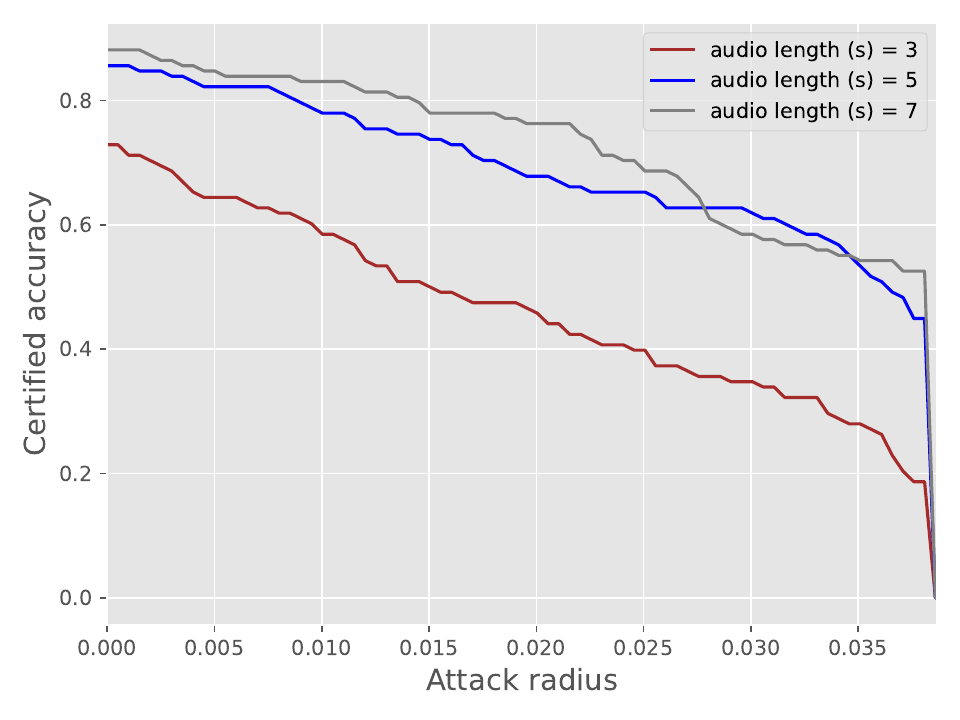}
         \caption{Dependency on the audio length $n$.}
     \end{subfigure}
        \caption{Pyannote model. Classification setting. Dependency of certified accuracy on $M$ and number of presented in enrollment set classes.}
        \label{fig:pyannote_clf_2}
\end{figure*}

\begin{figure*}[htb]
     \centering
     \begin{subfigure}{0.33\textwidth}
         \centering
         \includegraphics[width=\textwidth]{images/pyannote_SIGMA_other.pdf}
         \caption{Dependency on $\sigma$}
     \end{subfigure}
     \begin{subfigure}{0.33\textwidth}
         \centering
         \includegraphics[width=\textwidth]{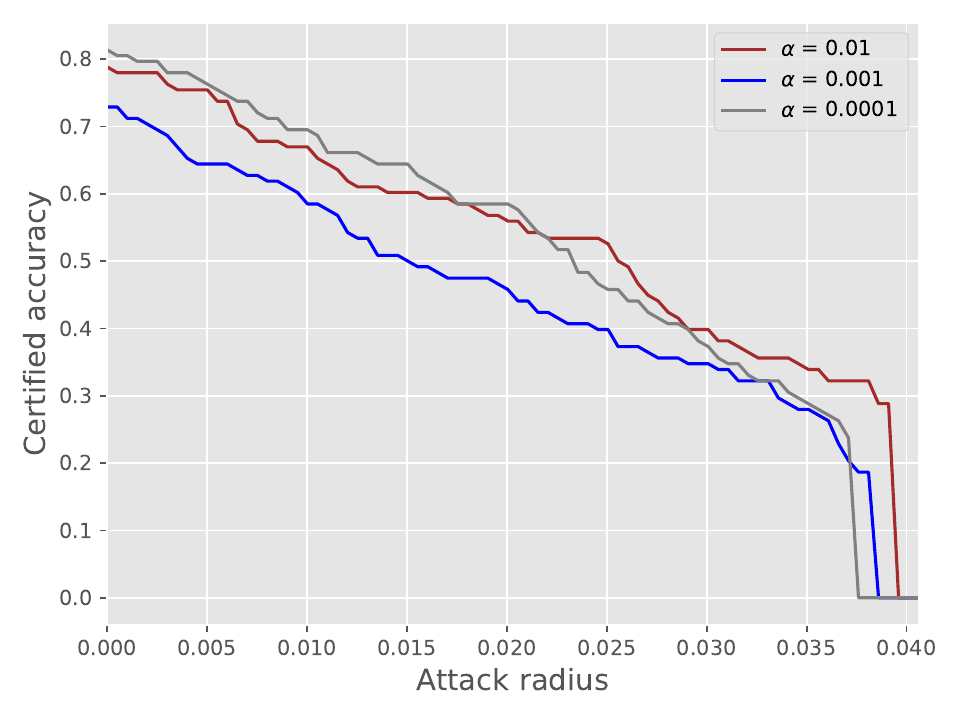}
         \caption{Dependency on $\alpha$}
     \end{subfigure}
     \begin{subfigure}{0.33\textwidth}
         \centering
         \includegraphics[width=\textwidth]{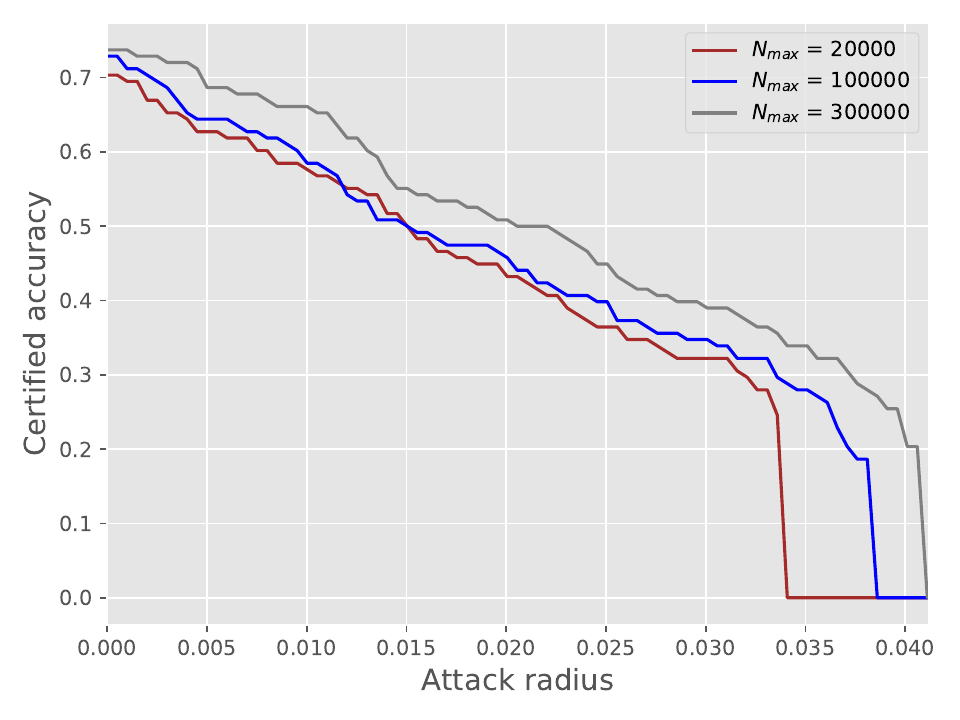}
         \caption{Dependency on $N_{\text{max}}$}
     \end{subfigure}
        \caption{Pyannote model. Classification setting. Dependency of certified accuracy on $\sigma$, $\alpha$, and $N_{\text{max}}$.}
        \label{fig:pyannote_clf_1}
\end{figure*}

\begin{figure*}[htb]
     \centering
     \begin{subfigure}{0.33\textwidth}
         \centering
         \includegraphics[width=\textwidth]{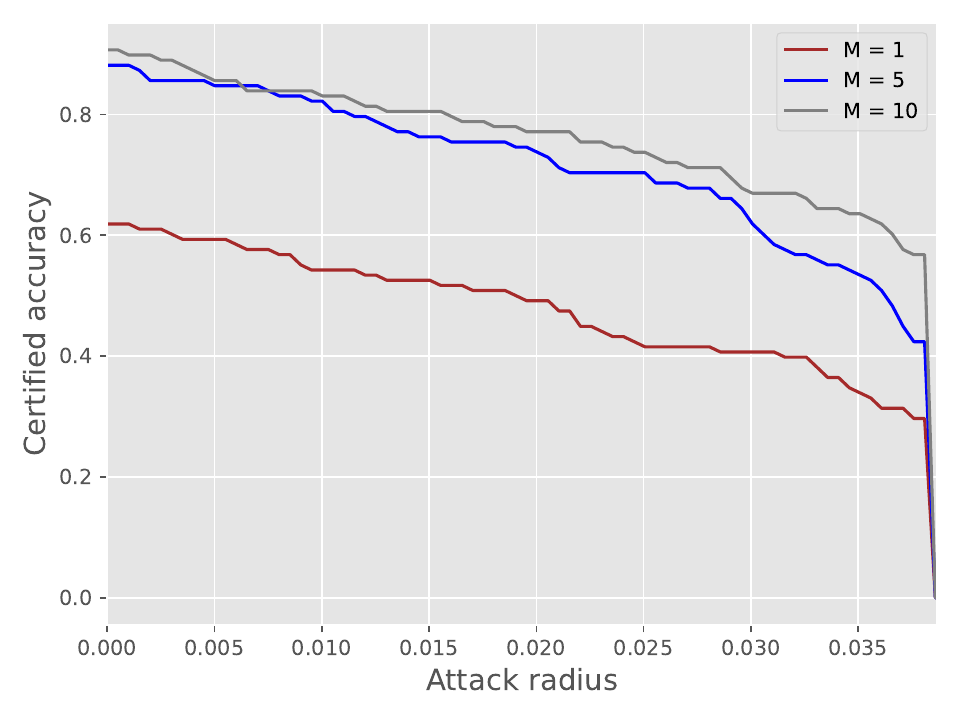}
         \caption{Dependency on $M$}
     \end{subfigure}
     \begin{subfigure}{0.33\textwidth}
         \centering
         \includegraphics[width=\textwidth]{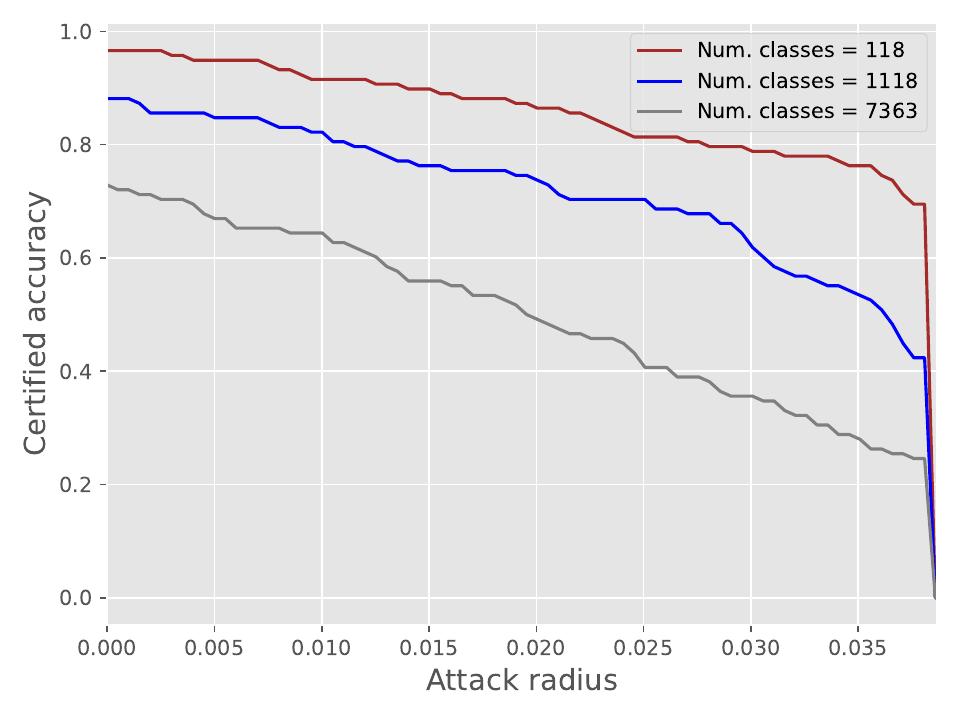}
         \caption{Dependency on number of speakers.}
     \end{subfigure}
     \begin{subfigure}{0.33\textwidth}
         \centering
         \includegraphics[width=\textwidth]{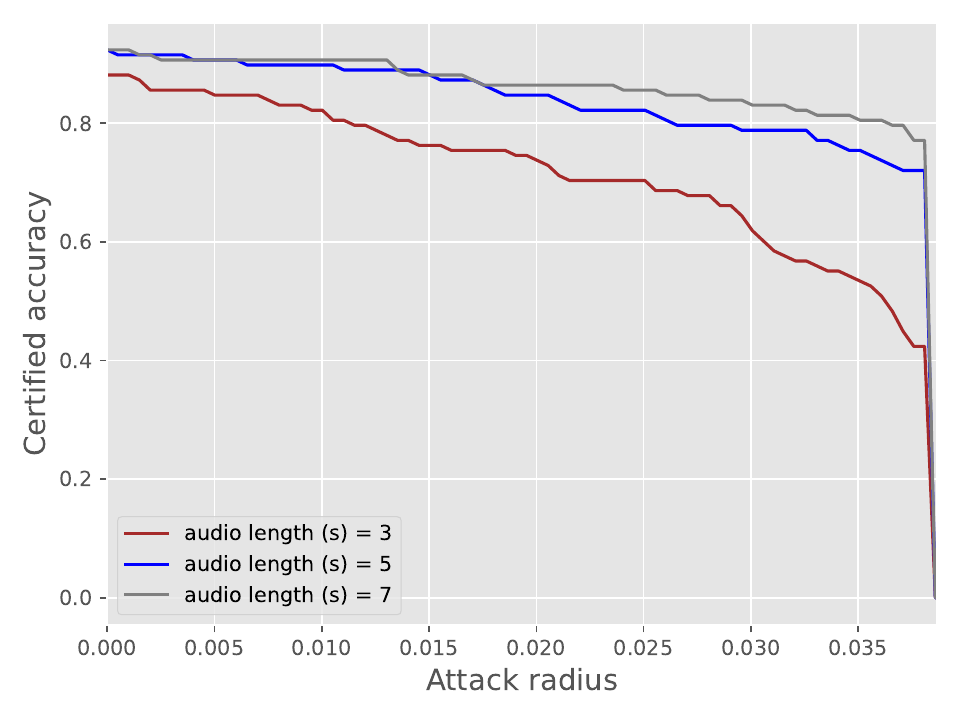}
         \caption{Dependency on the audio length.}
     \end{subfigure}
        \caption{ECAPA-TDNN model. Classification setting. Dependency of certified accuracy on $M$ and number of presented in enrollment set classes.}
        \label{fig:ecapa_clf_2}
\end{figure*}

\begin{figure*}[htb]
     \centering
     \begin{subfigure}{0.33\textwidth}
         \centering
         \includegraphics[width=\textwidth]{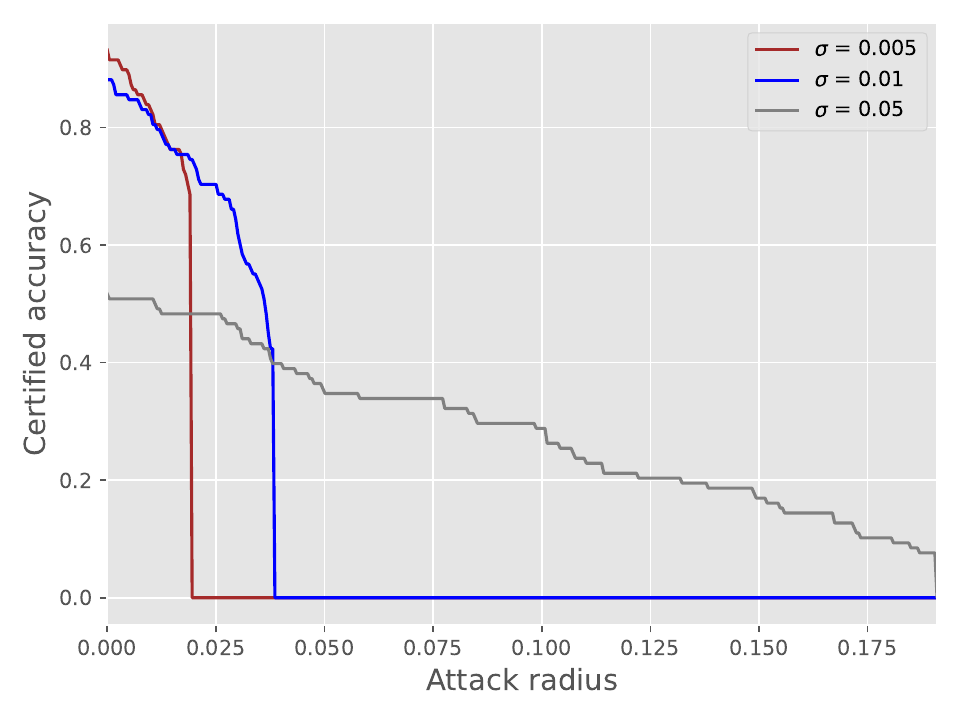}
         \caption{Dependency on $\sigma$}
     \end{subfigure}
     \begin{subfigure}{0.33\textwidth}
         \centering
         \includegraphics[width=\textwidth]{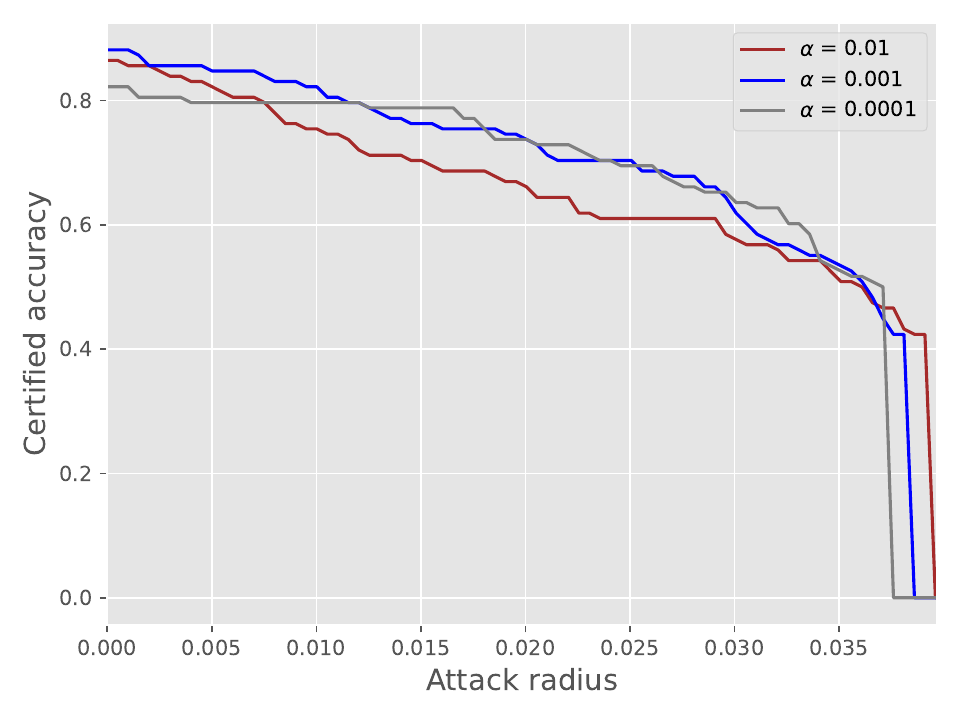}
         \caption{Dependency on $\alpha$}
     \end{subfigure}
     \begin{subfigure}{0.33\textwidth}
         \centering
         \includegraphics[width=\textwidth]{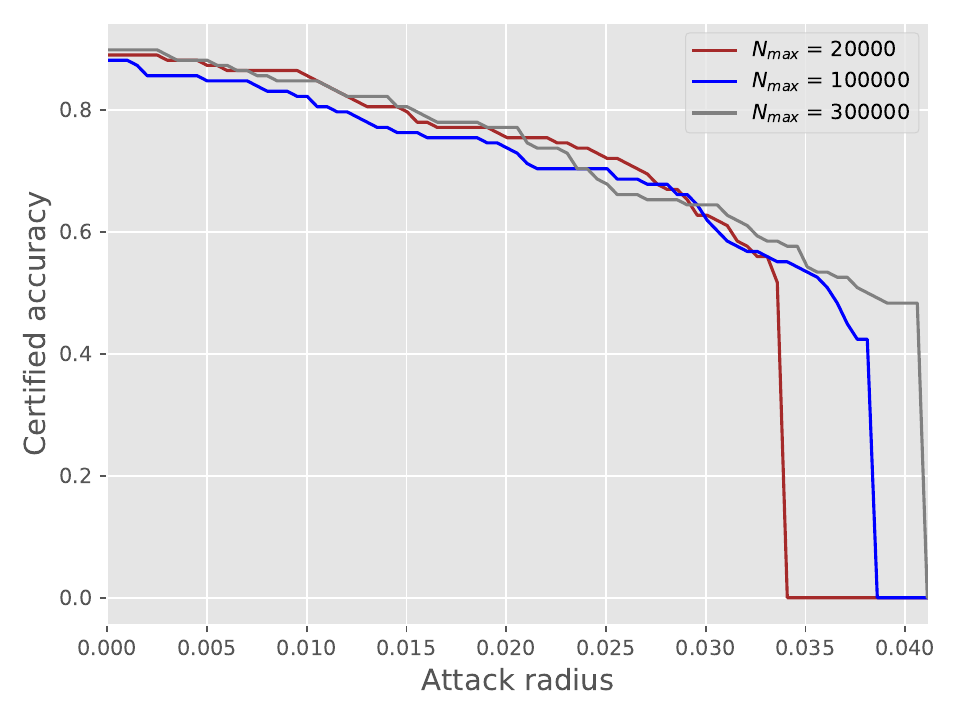}
         \caption{Dependency on $N_{\text{max}}$}
     \end{subfigure}
        \caption{ECAPA-TDNN model. Classification setting. Dependency of certified accuracy on $\sigma$, $\alpha$, and $N_{\text{max}}$.}
        \label{fig:ecapa_clf_1}
\end{figure*}


\end{document}